\documentclass[10pt,letterpaper]{article}
\usepackage[utf8]{inputenc}
\usepackage{amsmath}
\usepackage{amsfonts}
\usepackage{amssymb}
\usepackage[titletoc]{appendix}
\usepackage{authblk}
\usepackage{cite}
\usepackage{calc}
\usepackage{dsfont}
\usepackage{graphicx}
\usepackage{multirow}
\usepackage[left=2cm,right=2cm,top=2cm,bottom=2cm]{geometry}
\usepackage{subcaption}

\title{Generalized Uncertainty Principle and Angular Momentum}

\author{Pasquale Bosso\thanks{pasquale.bosso@uleth.ca} }
\author{Saurya Das\thanks{saurya.das@uleth.ca} }
\affil{Theoretical Physics Group and Quantum Alberta, \protect\\ Department of Physics and Astronomy, University of Lethbridge, \protect\\ 4401 University Drive, Lethbridge,\protect\\ Alberta, Canada, T1K 3M4}

\date{\vspace{-5ex}}

\begin{document}

\maketitle

\begin{abstract}
	Various models of quantum gravity suggest a modification of the Heisenberg's Uncertainty Principle, to the so-called Generalized Uncertainty Principle, between position and momentum.
	In this work we show how this modification influences the theory of angular momentum in Quantum Mechanics.
	In particular, we compute Planck scale corrections to angular momentum eigenvalues, the hydrogen atom spectrum, the Stern-Gerlach experiment and the Clebsch-Gordan coefficients.
	We also examine effects of the Generalized Uncertainty Principle on multi-particle systems.
\end{abstract}

\tableofcontents

\section{Introduction}

A successful formulation of a quantum theory of gravity, and its possible unification with the other fundamental interactions of Nature, remains as one of the foremost challenges of theoretical physics.
While various promising approaches for a Quantum Gravity (QG) theory have been proposed, such as Superstring Theory, Loop Quantum Gravity, etc., there has not been a single experiment or observation to our knowledge, which clearly supports or refutes any theory.
On the other hand, no experiment or observation has shown any deviation from the present theory of gravity.
Although this is normally attributed to the immensity of, and the related difficulty of directly accessing the Planck energy scale ($E_\mathrm{Pl} \sim 10^{16}$ TeV), there are indications that current experimental accuracies and those in the near future may be sufficient to detect such experimental signatures in at least some laboratory based experiments.

In the last couple of decades, it has been shown that most theories of QG predict momentum dependent modifications of the position-momentum commutation relation, the consequent modifications of the Heisenberg's Uncertainty Principle (HUP), and existence of a minimum measurable length near the Planck scale \cite{Amati1989_1,AmelinoCamelia2002_1,Garay1995_1,Gross1988_1,Maggiore1993_1,Maggiore1993_2,Scardigli1999_1}.
Studies of the so-called Generalized Uncertainty Principle (GUP) showed modifications in several areas of Quantum Mechanics (QM).
For example, modifications are expected for the Hamiltonian describing a minimal coupling with an electromagnetic field, as shown in \cite{Das2009_1}.
This result was then used to show modification of the Landau levels \cite{Das2009_1,Ali2011_1}.
It is possible to show that GUP affects also Lamb shift, the case of a potential step and of a potential barrier, a quantum configuration used in Scanning Tunnelling Microscopes \cite{Das2009_1,Ali2011_1}, the case of a particle in a box \cite{Ali2009_1}, showing that the length of the box is quantized when a GUP is considered, and the energy levels for a simple harmonic oscillator \cite{Ali2011_1}.
It has also been showed that GUP can also be detected considering its effects on quantum optical systems \cite{Pikovski2012_1}.

Carrying forward this program, in this article we compute QG corrections to an important theoretical as well as experimental area of quantum mechanics, namely the angular momentum of elementary quantum systems, and their experimental applications.
We indeed show that there are potential measurable Planck scale effects to well-understood phenomena such as line spectra from the hydrogen atom, the Stern-Gerlach experiment, Larmor frequency and Clebsh-Gordan coefficients.

Several models were proposed in the past years to account for this modification of the Heisenberg principle.
A quadratic model, for example, proposed in \cite{Maggiore1993_1,Scardigli1999_1,Kempf1995_1}, can account for this feature by adding a momentum dependent quadratic term in the position-momentum commutation relation.
In what follows, we will consider the most general commutation relation correct up to second order in momentum \cite{Ali2011_1}
\begin{equation}
	[x_i,p_j] = i \hbar \left[ \delta_{ij} - \alpha \left( p \delta_{ij} + \frac{p_i p_j}{p}\right) + \beta^2 ( p^2 \delta_{ij} + 3 p_i p_j) \right]~, \label{eqn:GUP}
\end{equation}
where $\alpha = \alpha_0 / M_P c$, $\beta = \beta_0 / M_P c$, $\alpha_0$ and $\beta_0$ are dimensionless parameters, normally assumed to be of order unity, and $M_P\sim 2\times10^{-8}$ Kg is the Planck Mass, while $i,j = 1,2,3$ represent different vector components.
One also assumes that position commutes with position, and momentum with momentum
\begin{equation}
	[x_i, x_j] = 0 = [p_i,p_j]~.
\end{equation}
This assumption implies the fulfillment of Jacobi identity, as shown in \cite{Ali2011_1}.
The Jacobi identity in the context of GUP was also studied in \cite{Pramanik2013_1}.
The quadratic model can be obtained by setting the linear term in \eqref{eqn:GUP} to zero.
In this work, we will consider the case $\alpha = \beta$.
The case $\alpha\not=\beta$ can be easily obtained following the same steps of this present paper.
In a one-dimensional problem, using the modified commutation relation, one can obtain the following inequality between uncertainties
\begin{equation}
	\begin{split}
		\Delta x \Delta p &\geq \frac{\hbar}{2}[1 - 2 \alpha \langle p \rangle  + 4 \beta^2 \langle p^2 \rangle] = \\
		&\geq \frac{\hbar}{2}\left[1 + \left( \frac{\alpha}{\sqrt{\langle p^2 \rangle}} + 4 \beta^2 \right) \Delta p^2 + 4 \beta^2 \langle p \rangle^2 - 2 \alpha \sqrt{\langle p^2 \rangle}\right]~.
	\end{split}
\end{equation}

Since the existence of a minimal measurable length implies the existence of a minimal angular resolution, we are forced to consider a GUP also for angle variables and their conjugate momenta, \emph{i.e.}, angular momenta.
We therefore expect a modification of the angular momentum algebra.
In the present paper we will obtain this modification as a consequence of the GUP in position and momentum.

This paper is organized as follows. In Section \ref{sec:algebra} we first present the modified angular momentum algebra and some useful relations for subsequent sections.
In Section \ref{sec:spectrum} we will show how the angular momentum spectrum is modified, illustrating some issues and assumptions made to solve them.
In Section \ref{sec:h-atom} we will apply our results to the hydrogen atom.
In Section \ref{sec:magnetic_field} we will consider magnetic interaction on an atom.
This interaction is relevant because, as described in Subsection \ref{subsec:stern-gerlach}, a direct observation of GUP effects in a Stern-Gerlach-like experiment may be possible.
In Section \ref{sec:multi-particles} we study systems composed of more than one particle, while in Subsection \ref{subsec:CG-coefficients} we introduce the modified Clebsch-Gordan coefficients in GUP.
Finally, in Section \ref{sec:conclusions} we summarize our results, list the open problems and discuss future directions.

\section{Modified Angular Momentum Algebra} \label{sec:algebra}

In this section we start with the standard definition of angular momentum in Classical Mechanics,
\begin{equation}
	\vec{L} = \vec{q} \times \vec{p} = \left(
	\begin{array}{c}
		yp_z - zp_y\\
		zp_x - xp_z\\
		xp_y - yp_x
	\end{array}
	\right)~. \label{eqn:def_L}
\end{equation}
GUP now implies the existence of minimal measurable angles.
Using the commutator (\ref{eqn:GUP}), one can show (see Appendix \ref{apx:commutator_LiLj}) that the commutation relation between angular momentum components is modified as 
\begin{equation}
	[L_i,L_j] = i \hbar \epsilon_{ijk} L_k (1 - \alpha p + \alpha^2 p^2)~. \label{eqn:generalized_commutator_ang_mom}
\end{equation}
It is also possible to show that Jacobi identity is trivially fulfilled by this modified commutation relation.
Note that we recover the standard commutation relation $[L_i,L_j] = i \hbar \epsilon_{ijk} L_k$ for $\alpha=0$.
Including GUP, one still has however
\begin{equation}
	[L^2,L_j] = 0~. \label{eqn:commutator_L2_Lj}
\end{equation}
Therefore, even with a modified angular momentum algebra, we can define simultaneous eigenstates of $L^2$ and $L_z$.
Furthermore we notice that, since for (\ref{eqn:commutator_Li_p2}) and (\ref{eqn:commutator_Li_p}) in Appendix \ref{apx:commutator_LiLj} also $p$ and $p^2$ commute with $L_z$, they also commute with $L^2$
\begin{align}
	[L^2,p] &= \sum_{i=1,2,3} L_i[L_i,p] + [L_i,p]L_i = 0~, & [L^2,p^2] &= \sum_{i=1,2,3} L_i[L_i,p^2] + [L_i,p^2]L_i = 0~.  \label{eqn:commutator_L2_p_p2}
\end{align}
Thus we can define simultaneous eigenstates of the operators $L^2$, $L_z$, and $p$, and moreover, such a state is also an energy eigenstate for the free case.

As proposed in \cite{Ali2011_1}, we will also expand the momentum operator $p$, subject to the modified commutation relation (\ref{eqn:GUP}), in terms of a low-energy momentum operator $p_0$, subject to the standard commutation relation, up to second order in $\alpha$.
The following relation is exact, such that (\ref{eqn:GUP}) is satisfied, as shown in \cite{Ali2011_1}
\footnote{As seen in Section \ref{sec:h-atom} of this paper, this relation poses no problem of analyticity for one effective spatial dimension, \emph{e.g.} $r$. For higher dimensions, the problem is circumvented by considering the Dirac equation \cite{Das2010_1,Deb2016_1}.}
\begin{equation}
	p_i = p_{0,i} ( 1 - \alpha p_0 + 2 \alpha^2 p_0^2), \qquad \mbox{where \bigskip} p_{0,i} = -i \hbar \frac{\partial}{\partial q_i}~. \label{eqn:henergy_lenergy}
\end{equation}
This expansion is sufficient to satisfy \eqref{eqn:GUP}.
It also gives us the necessary tool to compute GUP corrections to the quantum mechanical systems in what follows.
Notice that $\vec{p}_0$ corresponds to the generator of translations.
On the other hand, introducing the generators of rotations, one necessarily obtains
\begin{equation}
	[L_{0,i},L_{0,j}] = i \hbar \epsilon_{ijk} L_{0,k}~,
\end{equation}
and we can obtain an expansion of the physical angular momentum in terms of generators of translations and rotations
\begin{equation}
	L_i = L_{0,i} (1 - \alpha p_0 + 2 \alpha^2 p^2_0)~. \label{eqn:ang_mom_henergy_lenergy}
\end{equation}
Using this expansion, we find
\begin{equation}
		[L_i,L_j] = [L_{0,i},L_{0,j}](1 - \alpha p_0 + 2 \alpha^2 p^2_0)^2 = i \hbar \epsilon_{ijk} L_{0,k} (1 - 2 \alpha p_0 + 5 \alpha^2 p^2_0) = i \hbar \epsilon_{ijk} L_k (1 - \alpha p + \alpha^2 p^2)~. \label{eqn:com_rel_ang_mom_lenergy}
\end{equation}
We note that now the angular momentum does not coincide with the generator of rotations, and is therefore not conserved in general even for rotationally invariant Hamiltonians.
However, $\vec{L}$ can be expanded in terms of generators of rotations and translations.

\section{Modified Angular Momentum Spectrum}\label{sec:spectrum} \label{sec:angular_momentum}

As mentioned above, we consider simultaneous eigenstates
\begin{subequations}
	\begin{align}
		L^2|p\lambda m\rangle &= \hbar^2 \lambda |p\lambda m\rangle &  \lambda &\geq 0~,\label{eqn:eigenvalue_L2}\\
		L_z|p\lambda m\rangle &= \hbar m |p\lambda m\rangle & m^2 &\leq \lambda~. \label{eqn:eigenvalue_Lz}
	\end{align}
\end{subequations}
and define as usual
\begin{align}
	L_+ &= L_x + i L_y~, & L_- &= L_x - i L_y~. \label{def:L_pm}
\end{align}	
Using (\ref{eqn:generalized_commutator_ang_mom}), we get
\begin{subequations}
	\begin{align}
		[L_z,L_+] &= [L_z,L_x] + i [L_z,L_y] = \hbar (iL_y + L_x)(1 - \alpha p + \alpha^2 p^2) = \hbar L_+ (1 - \mathcal{C}) ~,\label{eqn:commutator_Lz_L+} \\
		[L_z,L_-] &= [L_z,L_x] - i [L_z,L_y] = \hbar (iL_y - L_x)(1 - \alpha p + \alpha^2 p^2) = - \hbar L_- (1 - \mathcal{C} )~, \label{eqn:commutator_Lz_L-}\\
		[L^2,L_+] &= [L^2,L_x] + i [L^2,L_y] = 0~, \label{eqn:commutator_L2_L+}\\
		[L^2,L_-] &= [L^2,L_x] - i [L^2,L_y] = 0~, \label{eqn:commutator_L2_L-}\\
		[L_+,L_-] &= [L_x,L_x] - i [L_x,L_y] + i[L_y,L_x] + [L_y,L_y] = -2i[L_x,L_y] = 2 \hbar L_z (1 - \mathcal{C} )~. \label{eqn:commutator_L+_L-}
	\end{align}
\end{subequations}
where $\mathcal{C} \equiv \alpha p - \alpha^2 p^2$ represents the modification due to the GUP.
From (\ref{eqn:commutator_L2_L+}, \ref{eqn:commutator_L2_L-}) we obtain
\begin{equation}
	L^2 L_\pm |p \lambda m\rangle = L_\pm L^2 | p \lambda m \rangle = \hbar^2 \lambda L_\pm |p \lambda m\rangle~.
\end{equation}
Thus, even for GUP, the ladder operators do not change the eigenvalues of the magnitude of the angular momentum.
On the other hand, using (\ref{eqn:commutator_Lz_L+}, \ref{eqn:commutator_Lz_L-}), we get
\begin{equation}
	L_z L_\pm |p \lambda m\rangle = L_\pm [ L_z \pm(1- \mathcal{C} )]|p \lambda m\rangle = \hbar [m \pm (1- \mathcal{C} )] L_\pm|p \lambda m\rangle~.
\end{equation}
That is, while $L_\pm$ still act as ladder operators, the spacing between two consecutive $L_z$ eigenstates undergoes modifications due to GUP.
Further, from (\ref{def:L_pm}) and (\ref{eqn:generalized_commutator_ang_mom}) we get
\begin{align}
	L_- L_+ &= L^2 - L_z[L_z + \hbar(1 - \mathcal{C} )]~, & L_+ L_- &= L^2 - L_z[L_z - \hbar(1 - \mathcal{C} )]~,
\end{align}
from which one gets the norms of the states $L_\pm |p \lambda m\rangle$ as
\begin{equation}
	||L_\pm | p \lambda m\rangle ||^2 = \langle p \lambda m | L_\mp L_\pm | p \lambda m \rangle = \hbar^2 \{ \lambda - m[m \pm (1- \mathcal{C} )]\} \geq 0~. \label{eqn:norm_L+-}
\end{equation}
We note that if we considered eigenstates of $L^2$ and $L_z$ alone, we would have obtained
\begin{equation}
	||L_\pm |\lambda m\rangle ||^2 =
	\langle \lambda m | L_\mp L_\pm | \lambda m \rangle =
	\hbar^2 \{ \lambda - m[m \pm (1-\langle \mathcal{C} \rangle)]\} \geq 0~,
\end{equation}
that is, the modification due to the GUP would appear as $\langle \cal{C} \rangle$.
This observation will come in handy when we will consider GUP effects for the hydrogen atom model in Sec. \ref{sec:h-atom}.
Also, notice that $1 - \langle \mathcal{C} \rangle$ is a positive quantity.
Indeed, since
\begin{equation}
	(\Delta p)^2 = \langle p^2 \rangle - \langle p \rangle^2 \geq 0 \qquad \Rightarrow \qquad \langle p^2 \rangle \geq \langle p \rangle^2 ~ ,
\end{equation}
one can write
\begin{equation}
	1 - \langle \mathcal{C} \rangle \geq \alpha^2 \langle p \rangle^2 - \alpha \langle p \rangle + 1 > 0 \qquad \forall \langle p \rangle \in \mathbb{R} ~ .
\end{equation}

Since $m^2 \leq \lambda$, there are upper and lower bounds on $m$, which we denote $m_+$ and $m_-$, respectively.
From (\ref{eqn:norm_L+-}), it follows
\begin{equation}
	L_\pm |p \lambda m_\pm\rangle = 0 \Rightarrow \lambda = m_\pm[m_\pm \pm (1- \mathcal{C} )]~. \label{eqn:condition_lambda_m}
\end{equation}
We can reach these values starting from any value $m$ and applying $s$ times $L_+$ or $t$ times $L_-$, with $s, t \in \mathbb{N}$
\begin{align}
	m_+ = & m + s(1- \mathcal{C} )~, & m_- = & m - t(1- \mathcal{C} )~.
\end{align}
Combining these two relations we find
\begin{equation}
	m_+ = m_- + (s+t)(1- \mathcal{C} ) = m_- + \eta(1- \mathcal{C} )~,
\end{equation}
with $\eta \in \mathbb{N}$ giving the distance between the two bounds.
Inserting this last relation in (\ref{eqn:condition_lambda_m}) and solving for $m_+$ or $m_-$ we find the two relations
\begin{align}
	m_+ = & \frac{\eta}{2}(1- \mathcal{C} ) = l (1- \mathcal{C} ), & m_- = & -\frac{\eta}{2}(1- \mathcal{C} ) = -l (1- \mathcal{C} )~, 
\end{align}
where we defined $l \equiv n/2$.
\begin{figure}
	\centering
	\begin{subfigure}[t]{0.45\linewidth}
		\includegraphics[width=\linewidth]{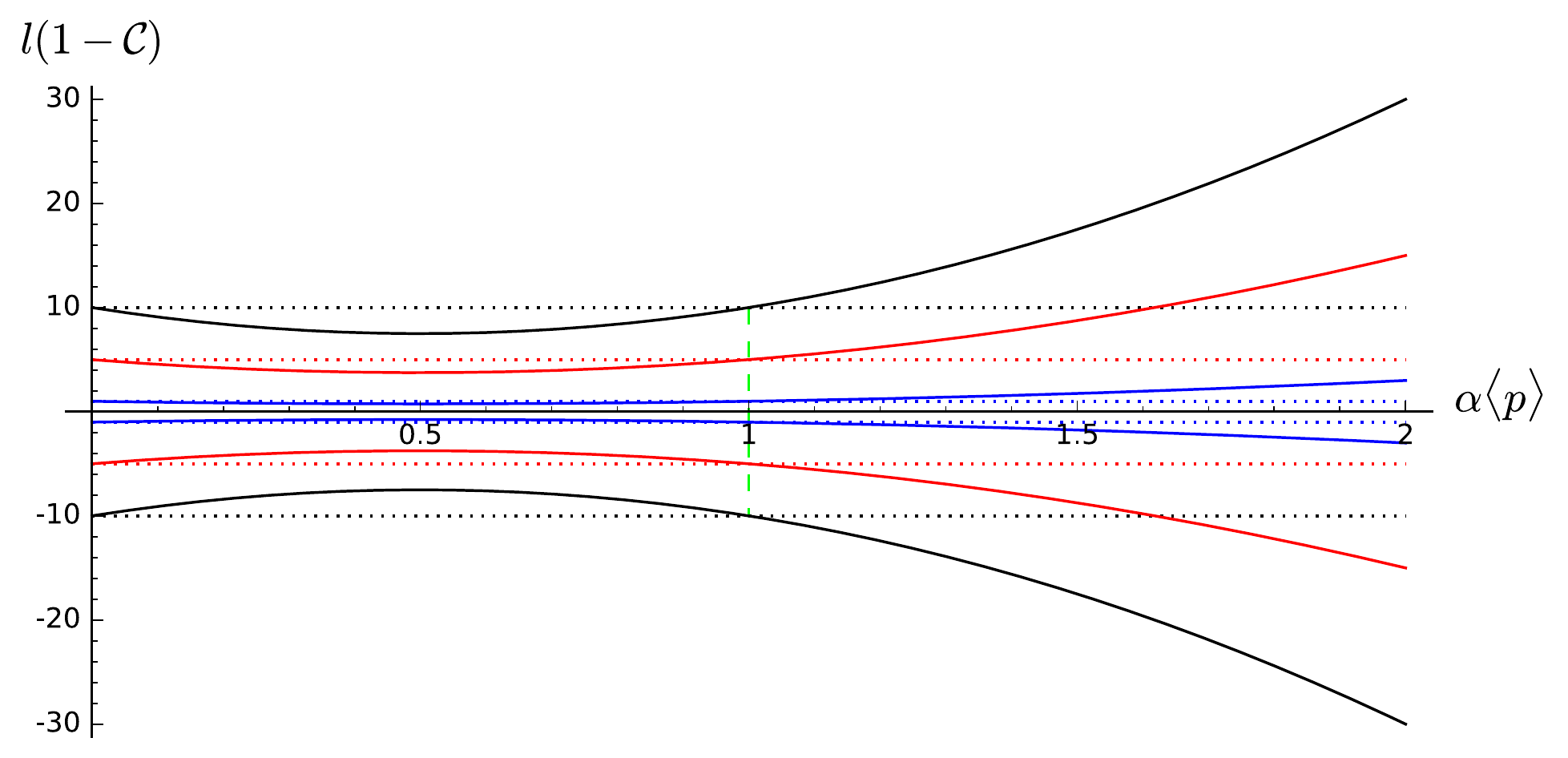}
    		\caption{Plot of the constraints with respect to $\alpha \langle p \rangle$ for three different values of the azimuthal quantum number. ${l=1}$, $l=5$ and $l=10$.} \label{fig:linear+quadratic}
    \end{subfigure}
    \qquad \qquad
    \begin{subfigure}[t]{0.45\linewidth}
		\includegraphics[width=\linewidth]{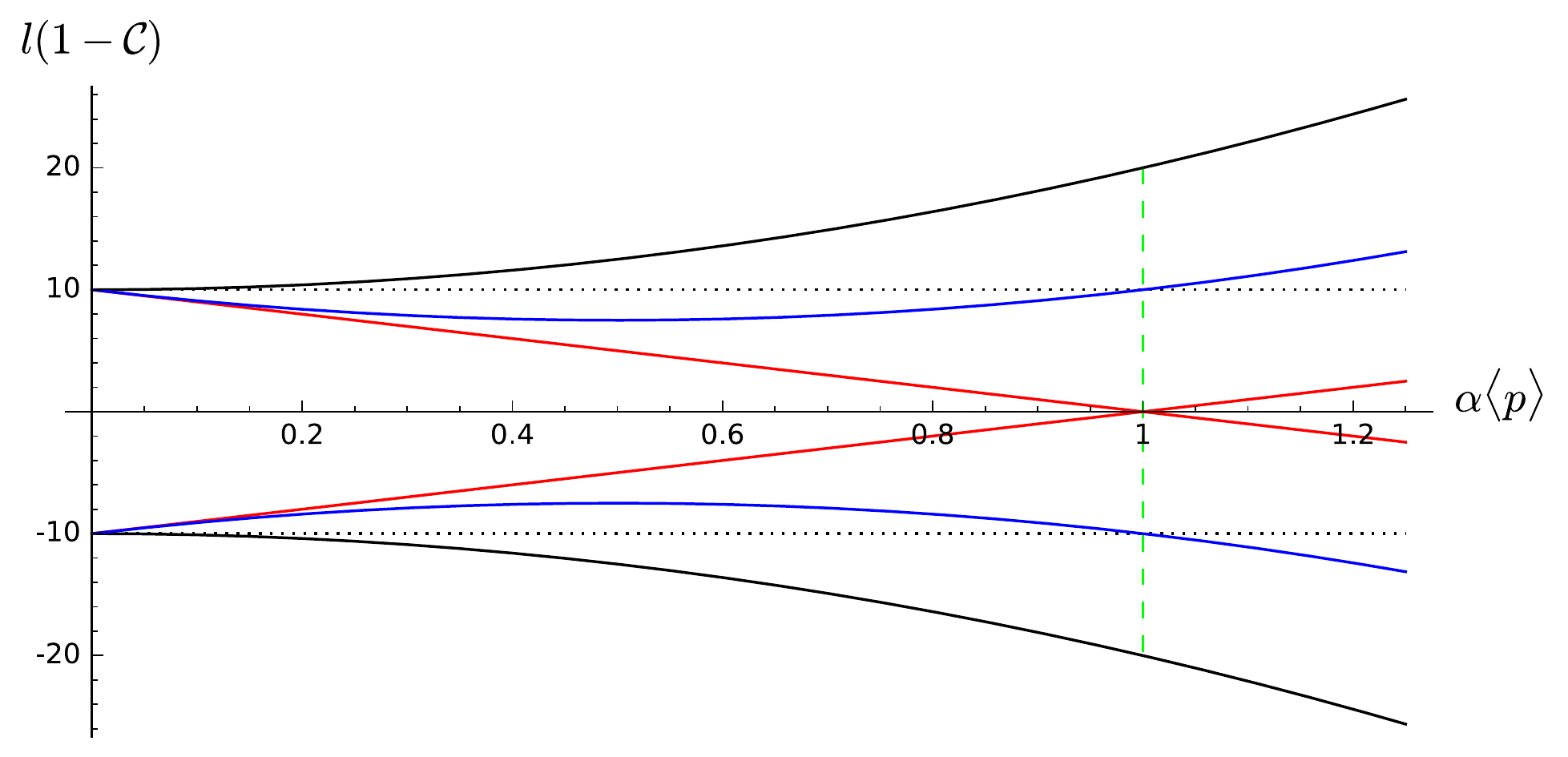}
    	\caption{Plot of the constraints with respect to $\alpha \langle p \rangle$ for $l=10$ considering three GUP models.}\label{fig:models}
	\end{subfigure}
\end{figure}
In Figure \ref{fig:linear+quadratic}, the bounds of $m$ are represented for a model including linear and quadratic terms for three values of $l$. 
Solid lines represent the constraints find in GUP, dotted lines the constraints in the standard theory.
In Figure \ref{fig:models}, the constraints for $l=10$ for three different parametrizations are plotted: a linear model (red line), a quadratic model (black line) and a linear and quadratic model (blue line).
In both the figures, the vertical dashed line corresponds to $\alpha \langle p \rangle = 1$.
Just for these plots, we assumed that $\alpha^2 \langle p^2 \rangle = (\alpha \langle p \rangle)^2$ for simplicity.

Using (\ref{eqn:condition_lambda_m}) we find
\begin{equation}
	\lambda = l(1- \mathcal{C} )[l(1- \mathcal{C} ) + (1- \mathcal{C} )] = l(l+1)(1- \mathcal{C} )^2~. 
\end{equation}
Notice that $m$ in general is not an integer and that two consecutive eigenstates of $L_z$, with eigenvalues $m_1$ and $m_2$, have $m_2 - m_1 = 1- \mathcal{C} $.
We can thus redefine the magnetic quantum number as
\begin{equation}
	m \rightarrow m(1- \mathcal{C} )~.
\end{equation}
This ``new'' $m$ is an integer bounded as follows
\begin{equation}
	-l \leq m \leq l~. \label{eqn:constraint_m}
\end{equation}
Next, for the eigenvalues equations (\ref{eqn:eigenvalue_L2}, \ref{eqn:eigenvalue_Lz}),
\begin{align}
	L^2|p l m\rangle &= \hbar^2 l(l+1) (1- \mathcal{C} )^2 |p l m\rangle~, & L_z|p l m\rangle &= \hbar m (1- \mathcal{C} ) |p l m\rangle~,\label{eqn:eigenvalues_L2_Lz_final}
\end{align}
Furthermore, for the $L_{\pm}$ we find
\begin{equation}
	L_{\pm}|plm\rangle = \hbar(1- \mathcal{C} )\sqrt{l(l+1) - m(m \pm 1)}|pl(m\pm1)\rangle~. \label{eqn:L+-}
\end{equation}

We can show that the eigenvalues of $L_z$ and $L^2$ summarized in (\ref{eqn:eigenvalues_L2_Lz_final}) are compatible with the uncertainty relation for angular momentum implied by (\ref{eqn:generalized_commutator_ang_mom}).
First, considering the definition of the operators $L_\pm$ in (\ref{def:L_pm}), we can show that the expectation values for $L_x$ and $L_y$ in an $L^2$ and $L_z$ eigenstate are zero
\begin{equation}
	\langle L_x \rangle  = \langle L_y \rangle = 0~,
\end{equation}
implying that the variance of these quantities for the same eigenstates are
\begin{equation}
	(\Delta L_x)^2 = \langle L_x^2 \rangle, \qquad (\Delta L_y)^2 = \langle L_y^2 \rangle~.
\end{equation}
Furthermore, for the second of (\ref{eqn:eigenvalues_L2_Lz_final}), the uncertainty relation for $L_x$ and $L_y$ on an $L_z$ eigenstate with eigenvalue $m \hbar$ is
\begin{equation}
	\Delta L_x \Delta L_y = (\Delta L_x)^2 = (\Delta L_y)^2 \geq \frac{\hbar}{2}|\langle L_z \rangle (1 - \langle \mathcal{C} \rangle) |= \frac{\hbar^2}{2} m (1 - \langle \mathcal{C} \rangle)^2~,
\end{equation}
where we used the equivalence between $L_x$ and $L_y$ in an $L_z$ eigenstate.
Inserting these last results in the first of (\ref{eqn:eigenvalues_L2_Lz_final}), we obtain
\begin{equation}
	\langle L^2 \rangle = \hbar^2 l(l+1) (1 - \langle \mathcal{C} \rangle)^2 = \langle L_x^2 \rangle + \langle L_y^2 \rangle + \langle L_z^2 \rangle \geq \hbar^2 m(m+1)(1 - \langle \mathcal{C} \rangle)^2~,
\end{equation}
where the equality holds for $m=l$.
We see that, as in standard QM, $\langle L_z^2 \rangle$ is bound by $\langle L^2 \rangle$ because of the uncertainty on the other components.

\section{Modified Energy Levels of the Hydrogen Atom} \label{sec:h-atom}
In this section we study GUP corrections to the hydrogen atom energy levels and spectra.
We start from the usual Hamiltonian for a central potential in electrostatic units \cite{Messiah}
\begin{equation}
	H \psi(\vec{r}) = \left[ \frac{p_r^2}{2 m} + \frac{L^2}{2 m  r^2} - \frac{e^2}{r} \right] \psi(\vec{r}) ~,
\end{equation}
where $p_r$ is understood to be the radial momentum.
As already noted in Sec. \ref{sec:angular_momentum}, considering an eigenstate of $L^2$ and $L_z$, GUP effects can be expressed in terms of the expectation value of $\langle \mathcal{C} \rangle = \alpha \langle p \rangle  - \alpha^2 \langle p^2 \rangle$.
In this section we will extend this observation, assuming that in general GUP modifications appear only through the expectation value $\langle \mathcal{C} \rangle$, \emph{i.e.}
\begin{align}
	L^2|l m\rangle &= \hbar^2 l(l+1) (1 - \langle\mathcal{C}\rangle )^2 |l m\rangle~, & L_z|l m\rangle &= \hbar m (1 - \langle\mathcal{C}\rangle ) |l m\rangle~, & \vec{p_r} & = \vec{p}_{0,r} (1 - \langle\mathcal{C}\rangle )~. \label{eqn:assumption_exp_value}
\end{align}
In this way, the correction term $\langle \mathcal{C} \rangle$ represents an average value for the GUP corrections.
This assumption can be further motivated noticing that, for experimental purposes, expectation values are the relevant quantities.
This suffices to estimate Planck scale corrections to observable quantities.
From (\ref{eqn:assumption_exp_value}) one can see that $\vec{L}$ is simply proportional to $\vec{L}_0$, and $L_i$ and $L^2$ commute with $H$.

With assumption (\ref{eqn:assumption_exp_value}), the GUP modified radial part of the Schr\"odinger equation is 
\begin{equation}
	\left[\frac{p_r^2}{2m} + \frac{\hbar^2 l(l+1)}{2m r^2}(1- \langle \mathcal{C} \rangle)^2 - \frac{e^2}{r}\right] y_l(r) =
	\left[\frac{p_{0,r}^2}{2m}(1 - \langle \mathcal{C} \rangle )^2 + \frac{\hbar^2 l(l+1)}{2m r^2}(1- \langle \mathcal{C} \rangle )^2 - \frac{e^2}{r}\right] y_l(r) = Ey_l(r)~. \label{eqn:radial_equation_1}
\end{equation}
where $E$ is the energy eigenstate.
Next, defining
\begin{equation}
\begin{array}{c}
	\displaystyle{\chi = \frac{\sqrt{-2mE}}{\hbar}~,
	\qquad a = \frac{\hbar^2}{m e^2}~,
	\qquad \nu = \frac{1}{a \chi}~, 
	\qquad \lambda = \frac{1}{1 - \langle \mathcal{C} \rangle}~,}\label{def:chi-a-nu} \\ \\
	\displaystyle{r \rightarrow z = 2 \lambda \chi r~,
	\qquad y_l (r) = z^{(l+1)} e^{-\frac{z}{2}} v(z)~,}
\end{array}
\end{equation}
the radial equation (\ref{eqn:radial_equation_1}) becomes
\begin{equation}
	\left[z\frac{d^2}{d \, z^2} + (2l + 2 - z)\frac{d}{d\,z} + \left(\frac{\nu}{1 - \langle \mathcal{C} \rangle }- l - 1\right)\right]v(z) = 0~, \label{eqn:laplace}
\end{equation}
which reduces to the well-known Laguerre equation for $\alpha=0$
\begin{equation}
	\left\{x\frac{d^2}{d \, x^2} + \left(2l+2 - x\right)\frac{d}{d\,x} + n'\right\}v(x) = 0~.
\end{equation}
Equation (\ref{eqn:laplace}), with the condition
\begin{equation}
	n' \equiv \frac{\nu}{1 - \langle \mathcal{C} \rangle } - l - 1 \in \mathbb{N}~, \label{eqn:def_and_condition_n'}
\end{equation}
has as solutions the associate Laguerre polynomials
\begin{equation}
	L^{(2l + 1)}_{n'} (z) = \sum_{i=0}^{n'} (-1)^i \frac{[(n' + 2l + 1)!]^2}{(n' - i)! (2l + 1 + i)!}\frac{z^i}{i!}~.
\end{equation}
Therefore, the solution for the radial Schr\"odinger's equation (\ref{eqn:radial_equation_1}) is
\begin{equation}
	y_l(r) = z^{(l+1)} e^{-\frac{z}{2}}\frac{n'! (2l + 1)!}{[(n' + 2l + 1)!]^1}L_{n'}^{(2l + 1)}(z) = 
	z^{(l+1)} e^{-\frac{z}{2}}\sum_{i=0}^{n'} (-1)^i \frac{n'! (2l + 1)!}{(n'-i)! (2l + 1 + i)}\frac{z^i}{i!}~.
\end{equation}
and the \emph{generalized principal quantum number} is
\begin{equation}
	n = \nu = \frac{e^2}{\hbar}\sqrt{\frac{m}{-2E}} = (n' + l + 1)(1 - \langle \mathcal{C} \rangle) = n_0 (1 - \langle \mathcal{C} \rangle)~,
\end{equation}
where $n_0$ is the principal quantum number of the standard theory.
In conclusion, the GUP modified energy levels of the hydrogen atom are given by
\begin{equation}
	E_n = - \frac{e^4}{\hbar^2}\frac{m}{2[(n' + l + 1)(1 - \langle \mathcal{C} \rangle)]^2} = -\left(\frac{e^2}{\hbar c}\right)^2 \frac{mc^2}{2n^2} \simeq E_{n(0)} [1 + 2 \alpha \langle p_0 \rangle + \alpha^2 (3 \langle p_0^2 \rangle - 4 \langle p_0 \rangle ^2) ]~, \label{eqn:energy_levels_GUP}
\end{equation}
where $E_{n(0)}$ is the corresponding energy level of the standard theory and where $\langle p_0 \rangle$ and $\langle p_0^2 \rangle$ are the expectation values of $p_0$ and $p_0^2$ in a $|n_0 l m\rangle$ eigenstate.
As before, we recover the results of the standard theory of the hydrogen atom for $\alpha=0$.

Eq. (\ref{eqn:energy_levels_GUP}) also implies that the wavenumber of photons emitted when the atom transits from an energy level $E_i$ to $E_f$ changes as follows
\begin{equation}
	\frac{1}{\lambda} = \frac{E_i - E_f}{h c} = R_\infty\left[\frac{1}{n_{0,f}^2(1 - \langle \mathcal{C}_f \rangle)^2} - \frac{1}{n_{0,i}^2(1 - \langle \mathcal{C}_i \rangle)^2}\right]~,
\end{equation}
where $R_\infty$ is the Rydberg constant.
We see now that the GUP-corrected spectrum depends not only on the principal quantum number, but also on the expectation values of the electron's momentum, as well as its angular momentum quantum numbers (the latter as we know also happens for the relativistic hydrogen atom).

\section{Inclusion of a Magnetic Field} \label{sec:magnetic_field}
Since a magnetic field interacts with the angular momentum of an atom via its magnetic moment, we expect that GUP modifications of the angular momentum theory will result in a modification of this interaction and have observable consequences.
In what follows we will consider atoms with only one electron in an S shell ($l=0$) and all other levels being filled.
Examples for these kind of atoms are those of the first group of the periodic table and the elements in the group of copper.
This may allow us to test the direct consequences of GUP on angular momenta studying its effects on the electronic spin.

We assume the spin operators to satisfy the same modified algebra for the angular momentum (\ref{eqn:generalized_commutator_ang_mom}) and the same ``low energy'' expansion (\ref{eqn:ang_mom_henergy_lenergy}).
The magnetic moment of an electron is
\begin{equation}
	\vec{M} = - \frac{g_S \mu_B}{\hbar}\vec{S}~,
\end{equation}
where 
\begin{equation}
	\mu_B = \frac{e \hbar}{2 m}~,
\end{equation}
is the Bohr magneton, $g_S$ is the electron g-factor and $\vec{S}$ is the spin operator.
Therefore we set 
\begin{equation}
	\vec{M} = \vec{M}_0(1 - \alpha p_0 + 2 \alpha^2 p_0^2)~, \label{eqn:mag_mom_henergy_lenergy}
\end{equation}
where
\begin{equation}
	\vec{M}_0 = - \frac{g_S \mu_B}{\hbar}\vec{S}_0~,
\end{equation}
satisfying the standard algebra, is interpreted as the magnetic moment at low energies.

For magnetic fields less than $\sim 10^6$ T \cite{Goswami}, the quadrupole term appearing in the Hamiltonian for the magnetic interaction on an atomic system is negligible with respect the dipole term.
Therefore, we can write the Hamiltonian considering only a term involving the scalar product between the magnetic moment and the magnetic field itself
\begin{equation}
	H = \frac{p^2}{2m} - \vec{M} \cdot \vec{B}~.
\end{equation}
Using (\ref{eqn:henergy_lenergy}) and (\ref{eqn:mag_mom_henergy_lenergy}), we can rewrite this relation in terms of low-energy quantities
\begin{equation}
	H = \frac{p^2_0}{2m}(1 - 2 \alpha p_0 + 5 \alpha^2 p_0^2) - (1 - \alpha p_0 + 2\alpha^2 p_0^2) \vec{M}_0\cdot\vec{B}~.
\end{equation}
Looking at the second term, \emph{i.e.} $(1 - \alpha p_0 + 2\alpha^2 p_0^2) \vec{M}_0\cdot\vec{B}$, we notice that it acts on both the space and the spin variables through the operators $p_0$ and $\vec{M}_0$ respectively.
As in the previous section, we replace $p_0$ and $p_0^2$ with $\langle p_0 \rangle$ and $\langle p_0^2 \rangle$, respectively.
In this way, the wavefunction can be factorized in its space and spin parts as follows
\begin{equation}
	\Psi(\vec{r},t) = \psi(\vec{r},t)[\alpha(t) | + \rangle + \beta(t) | - \rangle]~,
\end{equation}
with $\psi$ the spatial wave function, $\alpha$ and $\beta$ functions of time such that $|\alpha|^2 + |\beta|^2 = 1$, and $|+\rangle$ and $|-\rangle$ eigenstates of the $z$-component of the magnetic moment operator
\begin{align}
	M_z |+ \rangle &= \mu_0 |+\rangle~, & M_z |-\rangle &= - \mu_0 |-\rangle~.
\end{align}
Therefore the Schr\"odinger's equation also splits into
\begin{subequations}
	\begin{align}
		i\hbar \frac{\partial}{\partial t}\psi(\vec{r},t) &= \frac{p_0^2}{2m}( 1 - \langle \mathcal{C} \rangle )^2\psi(\vec{r},t)~,\\
		i\hbar \frac{d}{d\, t}(\alpha(t)|+\rangle + \beta(t)|-\rangle) &= - \vec{M}_0\cdot \vec{B}(1-\langle \mathcal{C} \rangle)(\alpha(t) | + \rangle + \beta(t) | - \rangle)~. \label{eqn:se_spin}
	\end{align}
\end{subequations}

We now show how this modification affects the magnetic interaction and how it could in principle be tested.

\subsection{Uniform Magnetic Field}
First we consider a uniform magnetic field along the $z$-axis, with the atom moving along the $y$-axis.
Eq. (\ref{eqn:se_spin}) for each component of the spinor can therefore be written as
\begin{subequations}
	\begin{align}
		i \hbar \frac{d}{d\,t}\alpha(t) &= - \mu_0 B(1- \langle \mathcal{C} \rangle)\alpha(t)~,\\
		i \hbar \frac{d}{d\,t}\beta(t) &= \mu_0 B(1- \langle \mathcal{C} \rangle)\beta(t)~,
	\end{align}
\end{subequations}
from which one can find the modified Larmor frequency of the system
\begin{equation}
	\omega_L = - \frac{2\mu_0B}{\hbar}(1- \langle \mathcal{C} \rangle)~. \label{eqn:larmor_frequency_GUP}
\end{equation}
Furthermore, since the magnitude of the linear momentum is not changed by the magnetic field, we can obtain the same form for the equations of motion as found in the standard theory
\begin{equation}
	\frac{d}{d\,t}\langle \vec{M} \rangle = \vec{\Omega} \times \langle \vec{M} \rangle~,
\end{equation}
where $\vec{\Omega} = \omega_L \hat{u}_z$.
We therefore see that the Larmor frequency is modified by the GUP, although the form of the precession equation remains unchanged.

\subsection{Non-Uniform Magnetic Field: Stern-Gerlach Experiment} \label{subsec:stern-gerlach}
In this subsection we will study the effects of GUP on the Stern-Gerlach experiment.

Consider a magnetic field with a gradient along the $z$-direction \cite{Basdevant}
\begin{equation}
	\vec{B}(\vec{r}) = B_z(\vec{r}) \hat{u}_z - b'x \hat{u}_x, \qquad \mbox{where} \qquad B_z(\vec{r}) = B_0 + b'z~,
\end{equation}
with $b'\Delta z \ll B_0$, where $\Delta z$ is the width of the beam used for the experiment along the $z$-axis.
The term $-b'x\hat{u}_x$ is necessary to ensure $\vec{\nabla}\cdot\vec{B}=0$.
If the dominant part of the field along $z$ is much more intense that the transverse component over the transverse extension of the wave packet, that is
\begin{equation}
	\langle M_z \rangle B_z \simeq \langle M_z \rangle B_0 \gg \langle M_x \rangle b' \Delta x~,
\end{equation}
then the eigenstates of $-\vec{M}\cdot\vec{B}$ remain practically equal to $|\pm\rangle_z$, since we can use the approximation $\vec{M} \cdot \vec{B} \simeq M_z B_z$, and we can neglect the transverse component.
For the original Stern-Gerlach experiment \cite{Bretislav2003_1} one has the following values
\begin{align}
	B_0 &\simeq 0.1\mbox{ T}~, & b' &\simeq 1\mbox{ T/mm}~, & \Delta z &\simeq \Delta x \simeq 0.03 \mbox{ mm}~,
\end{align}
that is
\begin{equation}
	\frac{B_0}{b' \Delta x} \simeq \frac{1}{0.3}~.
\end{equation}
With this assumption, the Schr\"odinger's equation for the two spin component is
\begin{equation}
	i \hbar \frac{\partial}{\partial t} \psi_\pm = \left[\frac{p_0^2}{2m} ( 1 - \langle \mathcal{C} \rangle )^2 \mp \mu_0(B_0 + b' z)(1- \langle \mathcal{C} \rangle )\right] \psi_\pm~.
\end{equation}

The expectation values of the position and of the momentum are given by
\begin{align}
	\langle \vec{r}_\pm\rangle &= \frac{\int \vec{r} |\psi_\pm(\vec{r},t)|^2 \mbox{d}^3r}{\int |\psi_\pm(\vec{r},t)|^2 \mbox{d}^3r}~, & \langle \vec{p}_\pm\rangle &= \frac{\int \psi_\pm^*(\vec{r},t) \vec{p}\psi_\pm(\vec{r},t) \mbox{d}^3r}{\int |\psi_\pm(\vec{r},t)|^2 \mbox{d}^3r} = \langle \vec{p}_{0\pm}\rangle (1 - \langle \mathcal{C} \rangle)~.
\end{align}
Consider the following set of equations derived from the Ehrenfest's Theorem
\begin{subequations}
	\begin{align}
		\frac{\mbox{d}}{\mbox{d} t}\langle \vec{r}_\pm\rangle &= \frac{\langle \vec{p}_\pm\rangle}{m} = \frac{\langle \vec{p}_{0\pm}\rangle}{m}(1 - \langle \mathcal{C} \rangle)~,\\
		\frac{\mbox{d}}{\mbox{d} t} \langle p_{x\pm}\rangle &= - \left\langle \mp \frac{\partial}{\partial x}\mu_0 B (1- \langle \mathcal{C} \rangle)\right\rangle = 0~,\\
		\frac{\mbox{d}}{\mbox{d} t} \langle p_{y\pm}\rangle &= - \left\langle \mp \frac{\partial}{\partial y}\mu_0 B (1-\langle \mathcal{C} \rangle)\right\rangle = 0~,\\
		\frac{\mbox{d}}{\mbox{d} t} \langle p_{z\pm}\rangle &= - \left\langle \mp \frac{\partial}{\partial z}\mu_0 B (1- \langle \mathcal{C} \rangle)\right\rangle = \pm \mu_0 b' (1- \langle \mathcal{C} \rangle)~.
	\end{align}
\end{subequations}
We assume that at $t=0$ a beam of atoms enters the apparatus, which we assume to be the origin of the coordinate system.
We also assume that the initial momentum is directed along the $y$-axis, without any component along the $x$ or $z$-axis, that is
\begin{subequations}
	\begin{align}
		\langle x_\pm \rangle(0) = \langle y_\pm \rangle(0) = \langle z_\pm \rangle(0) &= 0~,\\
		\langle p_{x\pm}\rangle(0) = \langle p_{z\pm}\rangle(0) &= 0~,\\
		\langle p_{y\pm}\rangle(0) &= mv~.
	\end{align}
\end{subequations}
We then obtain the following equations for its motion
\begin{subequations}
	\begin{align}
		\langle x_\pm \rangle(t) &= 0~,\\
		\langle y_\pm \rangle(t) &= vt~,\\
		\langle z_\pm \rangle(t) &= \pm \frac{\mu_0 b' t^2}{2m}(1- \langle \mathcal{C} \rangle)~.
	\end{align}
\end{subequations}
As for the standard theory, these equations represent a beam splitting in two along the $z$-axis.
In this case, though, the term $\langle \mathcal{C} \rangle$ will impose a dependence of the splitting on the expectation values of the momentum and the momentum squared of the electron in the outer S shell, the splitting being
\begin{equation}
	\delta z = \frac{\mu_0 b'}{m}\frac{L^2}{v^2}(1- \langle \mathcal{C} \rangle)~, \label{eqn:separation}
\end{equation}
where $L$ is the length of the apparatus.

Again, for the Stern-Gerlach apparatus described in \cite{Bretislav2003_1}, one has for an electron in the state 5S state of silver atoms
\begin{align}
	\langle p^2_0 \rangle &= 2.83\times10^{-26}\mbox{ N}^2\mbox{s}^2~, & \langle p_0 \rangle &= 0\mbox{ Ns}~.
\end{align}
We have that the ratio between the expected splitting in GUP and the splitting in the standard theory is
\begin{equation}
	\frac{\delta z_{\mathrm{GUP}}}{\delta z_0} - 1 = - \langle \mathcal{C} \rangle = \frac{\alpha_0^2 \langle p^2_0 \rangle}{(M_P c)^2} \simeq \alpha_0^2 6.66 \times 10^{-28}~. \label{eqn:cor_stern-gerlach}
\end{equation}
For splittings of the order achieved in the original experiment ($\sim 0.2$ mm), the difference between the GUP and the standard cases would not be observable.
But if larger splittings could be produced, for example with a longer apparatus or lower velocities of the atoms in the beam, a better resolution would be achieved.
On the other hand, one could also use atoms other than silver.
In the model used to describe the Stern-Gerlach experiment, though, a variation of the mass leads to two contrasting effects.
Consider for example, atoms heavier than silver. 
Since the inverse of the mass appears in (\ref{eqn:separation}), this will reduce the separation of the two spots.
On the other hand, higher atomic numbers lead to higher momenta for the external electrons, and hence to higher $|\langle \mathcal{C} \rangle|$.
The two effects will thus compete with each other.

\section{Multi-Particles Systems}\label{sec:multi-particles}

In this section we will examine how GUP affects multiparticle angular momentum algebra.

\subsection{Dependence of $[L_i,L_j]$ on the number of particles}

Consider a system of $N$ particles with angular momentum $\vec{l}_n$, with $n = 1, \ldots,N$, and the total angular momentum
\begin{equation}
	\vec{L} = \sum_{n=1}^N \vec{l}_n~.
\end{equation}
The commutator between components of the total angular momentum is
\begin{multline}
	[L_i,L_j] = \sum_{n=1}^N \sum_{m=1}^N[l_{i,n},l_{j,m}] = \sum_{n=1}^N[l_{i,n},l_{j,n}] = \sum_{n=1}^N i \hbar \epsilon_{ijk} l_{k,n} (1 - \alpha p_n + \alpha^2 p_n^2) \\
	= i \hbar \epsilon_{ijk} [L_k - \sum_{n=1}^N l_{k,n} ( \alpha p_n - \alpha^2 p_n^2)] =\\
	= i \hbar \epsilon_{ijk} \left[L_k (1 - \alpha P + \alpha^2 P^2) - \alpha \sum_{n=1}^N \left(l_{k,n} p_n - \frac{L_k P}{N}\right) + \alpha^2 \sum_{n=1}^N\left(l_{k,n} p_n^2 - \frac{L_k P^2}{N}\right)\right]~, \label{eqn:commutator_multi_particles}
\end{multline}
where
\begin{equation}
	P^2 = \sum_{n=1}^N \left[ p_n^2 + 2 \sum_{m>n}^{N-1} \vec{p}_n \cdot \vec{p}_m \right]~,
\end{equation}
and where we assumed
\begin{equation}
	[l_{i,m},l_{j,n}] = 0~, \qquad m \not = n~, \label{eqn:different_part_commutes}
\end{equation}
\emph{i.e.} the angular momentum components of different particles commute.

Consider for example the case in which all the particles in the system have the same angular momentum, \emph{e.g.} particle in a rotating ring,
\begin{equation}
	l_{n,k} = \frac{L_k}{N}~, \qquad n = 1, \ldots, N~,
\end{equation}
in which case from (\ref{eqn:commutator_multi_particles}), one obtains
\begin{equation}
	[L_i,L_j] = i \hbar \epsilon_{ijk} L_k \left[1 - \alpha P + \alpha^2 P^2 - \frac{\alpha}{N} \sum_{n=1}^N \left(p_n - P\right) + \frac{\alpha^2}{N} \sum_{n=1}^N\left(p_n^2 - P^2\right)\right]~. \label{eqn:ang_mom_com_rel_same_ang_mom}
\end{equation}

As a second example, we consider particles with the same linear momentum, \emph{e.g.}, a rigid body in pure translation, for which
\begin{equation}
	p_n = \frac{P}{N}~, \qquad n = 1,\ldots,N~,
\end{equation}
in which case we find
\begin{multline}
	[L_i,L_j] = i \hbar \epsilon_{ijk} \left[L_k (1 - \alpha P + \alpha^2 P^2) - \frac{\alpha}{N} P\sum_{n=1}^N \left(l_{k,n} - L_k\right) + \frac{\alpha^2}{N^2} P^2 \sum_{n=1}^N\left(l_{k,n} - N L_k\right)\right] = \\
	= i \hbar \epsilon_{ijk} L_k \left(1 - \frac{\alpha}{N} P + \frac{\alpha^2}{N^2} P^2\right)~. \label{eqn:ang_mom_com_rel_same_lin_mom}
\end{multline}
GUP in multiparticle system was also considered in \cite{Pramanik2014_1}.
Note that the RHS of (\ref{eqn:ang_mom_com_rel_same_ang_mom}) and (\ref{eqn:ang_mom_com_rel_same_lin_mom}) scale as different powers of $N$.

\subsection{Addition of Angular Momentum}

In this subsection we examine the problem of addition of angular momentum including GUP.

Consider a system composed by $N$ particles, with $l_n$ and $m_n$ the azimuthal and magnetic quantum numbers of the particles, with $n = 1, \ldots, N$.
From (\ref{eqn:eigenvalues_L2_Lz_final}) we have
\begin{subequations}
	\begin{align}
	l_n^2 |l_n,m_n\rangle &= \hbar^2 (1- \langle \mathcal{C}_n \rangle)^2l_n(l_n+1) |l_n,m_n\rangle~, \\
	l_{n,z} |l_n,m_n\rangle &= \hbar (1- \langle \mathcal{C}_n \rangle )m_n|l_n,m_n\rangle~.
	\end{align}
\end{subequations}
The $z$-component of the angular momentum for the composite system is
\begin{equation}
	L_z = \sum_{n=1}^N l_{n,z}~.
\end{equation}
Operating on the combined state
\begin{equation}
	|\{l_n,m_n\}\rangle  = \bigotimes_{n=1}^N |l_n,m_n\rangle~,
\end{equation}
we obtain
\begin{equation}
	L_z|\{l_n,m_n\}\rangle = \sum_{n=1}^N l_{n,z} |\{l_n,m_n\}\rangle = \sum_{n=1}^N\hbar(1- \langle \mathcal{C}_n \rangle)m_n|\{l_n,m_n\}\rangle~. \label{eqn:lz_composite_system}
\end{equation}
For the eigenvalue of $L_z$ in (\ref{eqn:lz_composite_system}) we thus have
\begin{multline}
	\sum_{n=1}^N\hbar(1- \langle \mathcal{C}_n \rangle )m_n = \sum_{n=1}^N\hbar m_n (1- \alpha \langle p_n \rangle + \alpha^2 \langle p_n^2 \rangle) = \\
	= \hbar \left[ M (1 - \alpha \langle P \rangle + \alpha^2 \langle P^2 \rangle) - \alpha \sum_{n=1}^N \left(m_n \langle p_n \rangle - \frac{M \langle P \rangle }{N}\right) + \alpha^2 \sum_{n=1}^N \left(m_n \langle p_n^2 \rangle - \frac{M \langle P^2 \rangle}{N}\right)\right]~,
\end{multline}
where $M = \sum m_n$.
Furthermore, for (\ref{eqn:constraint_m}) we have
\begin{equation}
	M_{\mathrm{min}} \equiv - \sum_{n=1}^N l_n \leq M \leq \sum_{n=1}^N l_n \equiv M_{\mathrm{max}}~,
\end{equation}
while, since the following inequalities hold
\begin{equation}
	-\sum_{n=1}^N l_n (1- \langle \mathcal{C}_n \rangle ) \leq \sum_{n=1}^N m_n (1- \langle \mathcal{C}_n \rangle ) \leq \sum_{n=1}^N l_n (1- \langle \mathcal{C}_n \rangle)~,
\end{equation}
we obtain for the eigenvalue of $L_z$
\begin{multline}
	M (1 - \langle \bar{\mathcal{C}} \rangle ) - \alpha \sum_{n=1}^N \left(m_n \langle p_n \rangle - \frac{M \langle P \rangle }{N}\right) + \alpha^2 \sum_{n=1}^N \left(m_n \langle p_n^2 \rangle - \frac{M \langle P^2\rangle }{N}\right) \geq \\
	\geq M_{\mathrm{min}} (1 - \langle \bar{\mathcal{C}}\rangle ) + \alpha \sum_{n=1}^N \left(l_n \langle p_n \rangle - \frac{M_{\mathrm{min}} \langle P \rangle }{N}\right) - \alpha^2 \sum_{n=1}^N \left(l_n \langle p_n^2 \rangle - \frac{M_\mathrm{min} \langle P^2 \rangle}{N}\right)
\end{multline}
and
\begin{multline}
	M (1 - \langle \bar{\mathcal{C}} \rangle ) - \alpha \sum_{n=1}^N \left(m_n \langle p_n \rangle - \frac{M \langle P \rangle }{N}\right) + \alpha^2 \sum_{n=1}^N \left(m_n \langle p_n^2 \rangle - \frac{M \langle P^2 \rangle }{N}\right) \leq \\
	\leq M_{\mathrm{max}} (1 - \langle \bar{\mathcal{C}} \rangle ) - \alpha \sum_{n=1}^N \left(l_n \langle p_n \rangle- \frac{M_{\mathrm{max}} \langle P \rangle}{N}\right) + \alpha^2 \sum_{n=1}^N \left(l_n \langle p_n^2 \rangle - \frac{M_\mathrm{max} \langle P^2 \rangle}{N}\right)~,
\end{multline}
where $\langle \bar{\mathcal{C}} \rangle = \alpha \langle P \rangle - \alpha^2 \langle P^2 \rangle $.

If $L$ is the azimuthal quantum number for the complete system, since $|M|\leq L$, we find that $L$ has the value
\begin{equation}
	L = \sum_{n=1}^N l_n~. \label{eqn:total_azimuthal_qn}
\end{equation}
Higher values of $L$ are not allowed, since they would imply $M > M_\mathrm{max}$.
Thus
\begin{equation}
	L_\mathrm{max} = \sum_{n=1}^N l_n
\end{equation}
or, for the case of two particles, useful for the next section,
\begin{equation}
	L_\mathrm{max} = l_1 + l_2~.\label{eqn:max_azimuthal_qn2}
\end{equation}
Note that the results (\ref{eqn:total_azimuthal_qn}-\ref{eqn:max_azimuthal_qn2}) are the same as in standard QM.
Therefore, following similar reasoning one gets
\begin{equation}
	|l_1 - l_2| \leq L \leq l_1 + l_2~.
\end{equation}

Next we define the ladder operators for the combined system
\begin{equation}
	L_{\pm} = \sum_{n=1}^N l_{n,\pm}~.
\end{equation}
Then it follows from for (\ref{eqn:commutator_Lz_L+}) and (\ref{eqn:commutator_Lz_L-}): 
\begin{equation}
	[l_{n,z},l_{m,\pm}] = \pm \delta_{nm} \hbar l_{n,\pm} ( 1 - \langle \mathcal{C}_n \rangle)~,
\end{equation}
from which one gets
\begin{multline}
	L_z L_\pm |l_1,m_1;\ldots;l_N,m_N\rangle = [L_\pm L_z \pm \hbar \sum_{n=1}^N l_{n,\pm} (1- \langle \mathcal{C}_n \rangle )] |l_1,m_1;\ldots;l_N,m_N\rangle = \\
	= \hbar \sum_{n=1}^N [M \pm (1- \langle \mathcal{C}_n \rangle)] l_{n,\pm} |l_1,m_1;\ldots;l_N,m_N\rangle~.
\end{multline}
Note that the RHS is no longer an eigenstate of $L_z$, unlike the $\alpha=0$ case.
Moreover, we get
\begin{equation}
	[L_+,L_-] = -i[L_x,L_y] +i[L_y,L_x] = -2i[L_x,L_y] = 2\hbar\sum_{n=1}^N l_{n,z}(1- \langle \mathcal{C}_n \rangle)~,
\end{equation}
where we have used (\ref{eqn:different_part_commutes}) and (\ref{eqn:generalized_commutator_ang_mom}).
Notice that this last commutator cannot be written in terms of the total angular momentum operator $L_z$ alone.

Next we specialize to the case of two angular momenta (\emph{i.e.} $N=2$)
\begin{equation}
	\begin{split}
		L_z L_\pm |l_1,m_1;l_2,m_2\rangle &= \hbar \{ [M \pm (1- \langle \mathcal{C}_1 \rangle)]l_{1,\pm} + [M \pm (1- \langle \mathcal{C}_2 \rangle)]l_{2,\pm} \}|l_1,m_1;l_2,m_2\rangle = \\
		&= \hbar \{ [M \pm (1- \langle \mathcal{C}_1 \rangle)]L_{\pm} \pm (\langle \mathcal{C}_1 \rangle - \langle \mathcal{C}_2 \rangle)l_{2,\pm} \}|l_1,m_1;l_2,m_2\rangle = \\
		&= \hbar \{ [M \pm (1- \langle \mathcal{C}_2 \rangle)]L_{\pm} \pm (\langle \mathcal{C}_2 \rangle -  \langle \mathcal{C}_1 \rangle)l_{1,\pm} \}|l_1,m_1;l_2,m_2\rangle~.
	\end{split}
\end{equation}
It is worth noticing that these equivalent results show that, not only we can obtain the results of standard QM by taking $\alpha=0$ but also when $\langle \mathcal{C}_1 \rangle = \langle \mathcal{C}_2 \rangle$.
We will find that this feature persists for the remainder of the section.

\subsection{Clebsch-Gordan Coefficients}\label{subsec:CG-coefficients}

As in standard QM, the following commutation relations still hold ($n=1,2$)
\begin{align}
	[L_i,l_n^2] & = 0 = [L^2,l_n^2]~, & [L_z,l_{n,z}] &= 0~,
\end{align}
but, in general
\begin{equation}
	[L^2,l_{n,z}]\not=0~.
\end{equation}
This means that $\{l_1^2,l_2^2,l_{1,z},l_{2,z}\}$ and $\{l_1^2,l_2^2,L^2,L_z\}$ form complete sets of observables also considering the GUP.
Since both the systems $\{|l_1,m_1;l_2,m_2 \rangle\}$ and $\{|l_1,l_2,L,M\rangle\}$ form complete sets of eigenstates, we use the completeness relations
\begin{align}
	\sum_{m_1,m_2}|l_1,m_1;l_2,m_2 \rangle \langle l_1,m_1;l_2,m_2| =& 1~, & \sum_{M,L} |l_1,l_2,L,M\rangle \langle l_1,l_2,L,M| =& 1~,
\end{align}
and write
\begin{subequations}
	\begin{align}
		|l_1,m_1;l_2,m_2 \rangle &= \sum_{M,L} |l_1,l_2,L,M\rangle \langle l_1,l_2,L,M|l_1,m_1;l_2,m_2 \rangle~,\label{eqn:definition_CG}\\
		|l_1,l_2,L,M\rangle &= \sum_{m_1,m_2}|l_1,m_1;l_2,m_2 \rangle \langle l_1,m_1;l_2,m_2|l_1,l_2,L,M\rangle~, \label{eqn:definition_CG_conjugate}
	\end{align}
\end{subequations}
where
\begin{equation}
	\langle l_1,m_1;l_2,m_2|l_1,l_2,L,M\rangle = \langle l_1,l_2,L,M|l_1,m_1;l_2,m_2 \rangle^*
\end{equation}
are the Clebsch-Gordan (CG) coefficients.

From here on, we use a simpler notation
\begin{align}
	|l_1,m_1;l_2,m_2 \rangle \equiv &  |m_1;m_2\rangle~, & \mbox{and} & & |l_1,l_2,L,M\rangle \equiv & |L,M\rangle~.
\end{align}

\subsubsection{Orthogonality relations}
Using the definition of CG coefficients in (\ref{eqn:definition_CG}) and (\ref{eqn:definition_CG_conjugate}) we can find the following orthogonality relations
\begin{subequations}
	\begin{align}
		\sum_{m1,m2} \langle l_1,l_2,L',M'|l_1,m_1;l_2,m_2 \rangle \langle l_1,m_1;l_2,m_2|l_1,l_2,L,M\rangle &= \delta_{M',M}\delta_{L',L}~, \label{eqn:orthogonality_relation_total}\\
		\sum_{M,L} \langle l_1,m_1';l_2,m_2'|l_1,l_2,L,M\rangle \langle l_1,l_2,L,M|l_1,m_1;l_2,m_2 \rangle &=\delta_{m_1',m_1}\delta_{m_2',m_2}~,\label{eqn:orthogonality_relation_parts}
	\end{align}
\end{subequations}
identical to those in standard QM.

\subsubsection{Clebsch-Gordan Recursion Relation}
From (\ref{eqn:L+-}) we have,
\begin{multline}
	L_\pm|L,M\rangle = \hbar(1- \langle \bar{\mathcal{C}} \rangle )\sqrt{L(L+1) - M(M\pm1)}|L,M\pm1\rangle = \\
	= \hbar(1- \langle \bar{\mathcal{C}} \rangle )\sqrt{L(L+1) - M(M\pm1)} \sum_{m_1,m_2}|m_1;m_2\rangle\langle m_1;m_2|L,M\pm1\rangle~,\label{eqn:rec_rel_step1}
\end{multline}
where $\langle P \rangle $ and $\langle P^2 \rangle$ are expectation values of the momentum and the momentum squared of the combined system.
On the other hand, since $L_\pm = l_{1,\pm} + l_{2,\pm}$, we have
\begin{multline}
	L_\pm|L,M\rangle = L_\pm\sum_{m_1,m_2}|m_1;m_2\rangle\langle m_1;m_2|L,M\rangle = \hbar \sum_{m_1,m_2}[(1- \langle \mathcal{C}_1 \rangle )\sqrt{l_1(l_1+1)-m_1(m_1\pm1)}|m_1 \pm 1;m_2\rangle + \\
	+ (1- \langle \mathcal{C}_2 \rangle )\sqrt{l_2(l_2+1)-m_2(m_2\pm1)}|m_1;m_2\pm1\rangle] \langle m_1;m_2|L,M\rangle = \\
	= \hbar \sum_{m_1,m_2}|m_1,m_2\rangle[(1- \langle \mathcal{C}_1 \rangle )\sqrt{l_1(l_1+1)-m_1(m_1\mp1)} \langle m_1\mp1;m_2|L,M\rangle + \\
	+ (1- \langle \mathcal{C}_2 \rangle )\sqrt{l_2(l_2+1)-m_2(m_2\mp1)}\langle m_1;m_2\mp1|L,M\rangle]~. \label{eqn:rec_rel_step2}
\end{multline}
Equating the RHS of (\ref{eqn:rec_rel_step1}) and (\ref{eqn:rec_rel_step2}) we get
\begin{multline}
	(1- \langle \bar{\mathcal{C}} \rangle )\sqrt{L(L+1) - M(M\pm1)}\langle m_1;m_2|L M\pm1\rangle = (1- \langle \mathcal{C}_1 \rangle )\sqrt{l_1(l_1+1)-m_1(m_1\mp1)} \langle m_1\mp1;m_2|L,M\rangle + \\
	+ (1- \langle \mathcal{C}_2 \rangle )\sqrt{l_2(l_2+1)-m_2(m_2\mp1)}\langle m_1;m_2\mp1|L,M\rangle~, \label{eqn:CG_recursive_relation}
\end{multline}
which reduces to the standard result if $\alpha = 0$, \emph{i.e.}, $\langle  \bar{\mathcal{C}} \rangle = \langle \mathcal{C}_1 \rangle=\langle \mathcal{C}_2 \rangle=0$.

\subsubsection{Clebsch-Gordan Coefficient Tables} \label{sssec:CG_tables}
In this section, using the results in Appendix \ref{apx:CG} we provide explicit expressions of  CG coefficients for some simple cases including the GUP.

\vspace{2em}

{\large $\mathbf{l_1 = 1/2, \quad l_2=1/2}$}

\vspace{1em}

$M=1$

\vspace{1 em}

\begin{tabular}{rc|c|}
	\cline{3-3}
	$L=$ && 1 \\
	\cline{2-3}
	\multicolumn{1}{c|}{$m_1,\quad m_2=$}&$1/2,\quad 1/2$ & 1 \\
	\cline{2-3}
\end{tabular}

\vspace{1em}

$M=0$

\vspace{1 em}

\begin{tabular}{rc|c|c|}
	\cline{3-4}
	$L=$ && 1 & 0\\
	\cline{2-4}
	\multicolumn{1}{c|}{\multirow{3}{*}{$m_1,\quad m_2=$}}&$1/2,\quad - 1/2$ & $\displaystyle{\frac{(1- \langle \mathcal{C}_2 \rangle )}{\sqrt{(1- \langle \mathcal{C}_1 \rangle )^2 + (1-\langle \mathcal{C}_2 \rangle )^2}}}$ & $ \displaystyle{\frac{(1- \langle \mathcal{C}_1 \rangle )}{\sqrt{(1- \langle \mathcal{C}_1 \rangle )^2 + (1- \langle \mathcal{C}_2 \rangle)^2}}}$ \\
	\cline{2-4}
	&\multicolumn{1}{|c|}{$-1/2, \quad 1/2$}&$\displaystyle{\frac{(1- \langle \mathcal{C}_1 \rangle )}{\sqrt{(1- \langle \mathcal{C}_1 \rangle)^2 + (1- \langle \mathcal{C}_2 \rangle )^2}}}$&$\displaystyle{- \frac{(1- \langle \mathcal{C}_2 \rangle ) }{\sqrt{(1- \langle \mathcal{C}_1 \rangle )^2 + (1 - \langle \mathcal{C}_2 \rangle)^2}}}$ \\
	\cline{2-4}
\end{tabular}

\vspace{2em}

{\large $\mathbf{l_1 = 1, \quad l_2=1/2}$}

\vspace{1em}

$M=3/2$

\vspace{1em}

\begin{tabular}{rc|c|}
	\cline{3-3}
	$L=$ && 3/2\\
	\cline{2-3}
	\multicolumn{1}{c|}{\multirow{1}{*}{$m_1,\quad m_2=$}}&$1,\quad 1/2$ & 1\\
	\cline{2-3}
\end{tabular}

\vspace{1em}

$M=1/2$

\vspace{1em}

\begin{tabular}{rc|c|c|}
	\cline{3-4}
	$L=$ && 3/2 & 1/2\\
	\cline{2-4}
	\multicolumn{1}{c|}{\multirow{3}{*}{$m_1,\quad m_2=$}}&$1,\quad - 1/2$ & $\displaystyle{\frac{(1- \langle \mathcal{C}_2 \rangle )}{\sqrt{2(1- \langle \mathcal{C}_1 \rangle )^2 + (1- \langle \mathcal{C}_2 \rangle )^2}}}$ & $ \displaystyle{\frac{\sqrt{2}(1- \langle \mathcal{C}_1 \rangle )}{\sqrt{2(1- \langle \mathcal{C}_1 \rangle )^2 + (1- \langle \mathcal{C}_2 \rangle )^2}}}$ \\
	\cline{2-4}
	&\multicolumn{1}{|c|}{$0, \quad 1/2$}&$\displaystyle{\frac{\sqrt{2}(1- \langle \mathcal{C}_1 \rangle )}{\sqrt{2(1- \langle \mathcal{C}_1 \rangle )^2 + (1- \langle \mathcal{C}_2 \rangle )^2}}}$&$\displaystyle{- \frac{(1- \langle \mathcal{C}_2 \rangle ) }{\sqrt{2(1- \langle \mathcal{C}_1 \rangle )^2 + (1- \langle \mathcal{C}_2 \rangle )^2}}}$ \\
	\cline{2-4}
\end{tabular}

\vspace{1em}

$M=-1/2$

\vspace{1em}

\begin{tabular}{rc|c|c|}
	\cline{3-4}
	$L=$ && 3/2 & 1/2\\
	\cline{2-4}
	\multicolumn{1}{c|}{\multirow{3}{*}{$m_1,\quad m_2=$}}&$0,\quad - 1/2$ & $\displaystyle{\frac{\sqrt{2}(1- \langle \mathcal{C}_2 \rangle )}{\sqrt{(1- \langle \mathcal{C}_1 \rangle )^2 + 2(1- \langle \mathcal{C}_2 \rangle )^2}}}$ & $ \displaystyle{\frac{(1- \langle \mathcal{C}_1 \rangle )}{\sqrt{(1- \langle \mathcal{C}_1 \rangle )^2 + 2(1- \langle \mathcal{C}_2 \rangle )^2}}}$ \\
	\cline{2-4}
	&\multicolumn{1}{|c|}{$-1, \quad 1/2$}&$\displaystyle{\frac{(1- \langle \mathcal{C}_1 \rangle )}{\sqrt{(1- \langle \mathcal{C}_1 \rangle )^2 + 2(1- \langle \mathcal{C}_2 \rangle )^2}}}$&$\displaystyle{- \frac{\sqrt{2}(1- \langle \mathcal{C}_2 \rangle ) }{\sqrt{(1- \langle \mathcal{C}_1 \rangle )^2 + 2(1- \langle \mathcal{C}_2 \rangle )^2}}}$ \\
	\cline{2-4}
\end{tabular}

\vspace{2em}

{\large $\mathbf{l_1=1, \quad l_2=1}$}

\vspace{1em}

$M=2$

\vspace{1em}

\begin{tabular}{rc|c|}
	\cline{3-3}
	$L=$ && 2\\
	\cline{2-3}
	\multicolumn{1}{c|}{\multirow{1}{*}{$m_1,\quad m_2=$}}&$1,\quad 1$ & 1 \\
	\cline{2-3}
\end{tabular}

\vspace{1em}

$M=1$

\vspace{1em}

\begin{tabular}{rc|c|c|}
	\cline{3-4}
	$L=$ && 2 & 1\\
	\cline{2-4}
	\multicolumn{1}{c|}{\multirow{3}{*}{$m_1,\quad m_2=$}}&$1,\quad 0$ & $\displaystyle{\frac{(1- \langle \mathcal{C}_2 \rangle )}{\sqrt{(1- \langle \mathcal{C}_1 \rangle )^2 + (1- \langle \mathcal{C}_2 \rangle )^2}}}$ & $ \displaystyle{\frac{(1- \langle \mathcal{C}_1 \rangle )}{\sqrt{(1- \langle \mathcal{C}_1 \rangle )^2 + (1- \langle \mathcal{C}_2 \rangle )^2}}}$ \\
	\cline{2-4}
	&\multicolumn{1}{|c|}{$0, \quad 1$}&$\displaystyle{\frac{(1- \langle \mathcal{C}_1 \rangle )}{\sqrt{(1- \langle \mathcal{C}_1 \rangle )^2 + (1- \langle \mathcal{C}_2 \rangle )^2}}}$&$\displaystyle{- \frac{(1- \langle \mathcal{C}_2 \rangle ) }{\sqrt{(1- \langle \mathcal{C}_1 \rangle )^2 + (1- \langle \mathcal{C}_2 \rangle )^2}}}$ \\
	\cline{2-4}
\end{tabular}

\vspace{1em}

$M=0$

\vspace{1em}

\begin{tabular}{rc|c|}
	\cline{3-3}
	$L=$ && 2 \\
	\cline{2-3}
	\multicolumn{1}{c|}{\multirow{5}{*}{$m_1$, $m_2=$}}&$1$, $- 1$ & $\displaystyle{\frac{(1- \langle \mathcal{C}_2 \rangle )^2}{\sqrt{[(1- \langle \mathcal{C}_1 \rangle )^2 + (1- \langle \mathcal{C}_2 \rangle )^2]^2 + 2(1- \langle \mathcal{C}_1 \rangle )^2(1- \langle \mathcal{C}_2 \rangle )^2}}}$ \\
	\cline{2-3}
	&\multicolumn{1}{|c|}{$0, \quad 0$} & $\displaystyle{\frac{2(1- \langle \mathcal{C}_1 \rangle )(1- \langle \mathcal{C}_2 \rangle )}{\sqrt{[(1- \langle \mathcal{C}_1 \rangle )^2 + (1- \langle \mathcal{C}_2 \rangle )^2]^2 + 2(1- \langle \mathcal{C}_1 \rangle )^2(1- \langle \mathcal{C}_2 \rangle )^2}}}$ \\
	\cline{2-3}
	&\multicolumn{1}{|c|}{$-1, \quad 1$} & $\displaystyle{\frac{(1- \langle \mathcal{C}_1 \rangle )^2}{\sqrt{[(1- \langle \mathcal{C}_1 \rangle )^2 + (1- \langle \mathcal{C}_2 \rangle )^2]^2 + 2(1- \langle \mathcal{C}_1 \rangle )^2(1- \langle \mathcal{C}_2 \rangle )^2}}}$ \\
	\cline{2-3}
\end{tabular}

\vspace{1em}

\begin{tabular}{rc|c|}
	\cline{3-3}
	$L=$ && 1\\
	\cline{2-3}
	\multicolumn{1}{c|}{\multirow{5}{*}{$m_1$, $m_2=$}}&$1$, $- 1$ & $\displaystyle{\frac{(1- \langle \mathcal{C}_1 \rangle )(1- \langle \mathcal{C}_2 \rangle )}{\sqrt{(1- \langle \mathcal{C}_1 \rangle )^4 + (1- \langle \mathcal{C}_2 \rangle )^4}}}$ \\
	\cline{2-3}
	&\multicolumn{1}{|c|}{$0, \quad 0$} & $\displaystyle{\frac{(1-\langle \mathcal{C}_1 \rangle)^2 - (1-\langle \mathcal{C}_2 \rangle)^2}{\sqrt{(1-\langle \mathcal{C}_1 \rangle)^4 + (1-\langle \mathcal{C}_2 \rangle)^4}}}$\\
	\cline{2-3}
	&\multicolumn{1}{|c|}{$-1, \quad 1$} & $\displaystyle{-\frac{(1-\langle \mathcal{C}_1 \rangle)(1-\langle \mathcal{C}_2 \rangle)}{\sqrt{(1-\langle \mathcal{C}_1 \rangle)^4 + (1-\langle \mathcal{C}_2 \rangle)^4}}}$ \\
	\cline{2-3}
\end{tabular}

\vspace{1em}

\begin{tabular}{rc|c|}
	\cline{3-3}
	$L=$ && 0 \\
	\cline{2-3}
	\multicolumn{1}{c|}{\multirow{5}{*}{$m_1$, $m_2=$}}&$1$, $- 1$ & $\displaystyle{\frac{2(1- \langle \mathcal{C}_1 \rangle )(1- \langle \mathcal{C}_2 \rangle )}{\sqrt{[(1- \langle \mathcal{C}_1 \rangle )^2 + (1- \langle \mathcal{C}_2 \langle )^2]^2 + 8(1- \langle \mathcal{C}_1 \rangle )(1- \langle \mathcal{C}_2 \rangle )}}} $ \\
	\cline{2-3}
	&\multicolumn{1}{|c|}{$0, \quad 0$} & $\displaystyle{-\frac{(1-\langle \mathcal{C}_1 \rangle)^2 + (1-\langle \mathcal{C}_2 \rangle)^2}{\sqrt{[(1-\langle \mathcal{C}_1 \rangle)^2 + (1-\langle \mathcal{C}_2 \rangle)^2]^2 + 8(1-\langle \mathcal{C}_1 \rangle)(1-\langle \mathcal{C}_2 \rangle)}}}$ \\
	\cline{2-3}
	&\multicolumn{1}{|c|}{$-1, \quad 1$} & $\displaystyle{\frac{2(1-\langle \mathcal{C}_1 \rangle)(1-\langle \mathcal{C}_2 \rangle)}{\sqrt{[(1-\langle \mathcal{C}_1 \rangle)^2 + (1-\langle \mathcal{C}_2 \rangle)^2]^2 + 8(1-\langle \mathcal{C}_1 \rangle)(1-\langle \mathcal{C}_2 \rangle)}}}$ \\
	\cline{2-3}
\end{tabular}\\
Notice that these tables reduce to the corresponding CG tables of standard QM, as given \emph{e.g.} in \cite{Goswami}, when $\alpha=0$ or when $\langle \mathcal{C}_1\rangle = \langle \mathcal{C}_2 \rangle$.

Note that the CG that we collected in these tables are found starting from a state of maximum total angular momentum and applying $L_-$ or applying orthonormality conditions with states previously analyzed.
Inverting the direction of our derivation, though, leads to slightly different coefficients.
In particular the two modifications $\langle \mathcal{C}_1\rangle$ and $\langle \mathcal{C}_2 \rangle$ are inverted.
Consider, for example, a system composed of two $l=1/2$ particles and total angular momentum $L=1$ and $M=0$.
Using the corresponding table we find
\begin{equation}
	\left\langle \frac{1}{2}; - \frac{1}{2} \middle| 1, 0 \right\rangle = \frac{(1- \langle \mathcal{C}_2 \rangle )}{\sqrt{(1- \langle \mathcal{C}_1 \rangle )^2 + (1- \langle \mathcal{C}_2 \rangle)^2}}~.
\end{equation}
On the other hand, starting from the state $|1,-1\rangle$ and applying the operator $L_+$, we find for the same CG
\begin{equation}
	\left\langle \frac{1}{2}; - \frac{1}{2} \middle| 1, 0 \right\rangle = \frac{(1- \langle \mathcal{C}_1 \rangle )}{\sqrt{(1- \langle \mathcal{C}_1 \rangle )^2 + (1- \langle \mathcal{C}_2 \rangle)^2}}~.
\end{equation}
The same result can be obtained considering a linear GUP model.
Ultimately this ambiguity is generated by the commutation relations (\ref{eqn:commutator_Lz_L+} - \ref{eqn:commutator_Lz_L-}).
These relations are in turn directly derived from (\ref{eqn:generalized_commutator_ang_mom}).
They are therefore implications of the modification of Heisenberg algebra and of the classical definition of angular momentum.
Further work is required for a better understanding of this ambiguity.

\section{Conclusions} \label{sec:conclusions}
A large number of theories predict a modification of the Heisenberg's Uncertainty Principle \cite{Amati1989_1,AmelinoCamelia2002_1,Garay1995_1,Gross1988_1,Maggiore1993_1,Maggiore1993_2,Scardigli1999_1}, where the commutation relation between canonical coordinates and momenta is modified by terms depending on the momentum.
Motivated by the possibility of being able to measure Planck scale effects in low energy quantum systems (as demonstrated earlier for various other examples), in this paper we have considered in detail the effect of GUP on angular momentum in QM.
Starting with the most general linear $+$ quadratic GUP, we first compute the corrected angular momentum algebra.
From this, we found the modified spectrum of the angular momentum operators $L_z$ and $L^2$.
We then found that these modifications lead to corrected energy levels of the hydrogen atom, and its behavior in an external magnetic field.
When applied to magnetic field interaction, it leads to different values for the Larmor frequency and for the splitting in the Stern-Gerlach experiment.
We finally showed how the modified algebra of the total angular momentum of a multi-particle system depends on the number of components and interesting Planck scale modifications to the CG coefficients.
It is worth noting that all the modifications derived in this paper are potentially observable allowing new tests on quantum gravity phenomenology.

It is worthwhile to note that in this paper we compute GUP corrections of the angular momentum spectrum and applied it to the appropriate Schr\"odinger equation governing non-relativistic systems, such as the hydrogen atom and the Stern-Gerlach experiment.
    While the corrections to the spectrum will equally apply to relativistic systems (such as relativistic hydrogen atom), to compute GUP corrections to energy eigenvalues and eigenvectors, the emission spectra etc., one would have to use Dirac equation.

Comparing the results of the present paper, we can obtain upper bounds for the GUP parameter $\alpha_0$.
For instance, if one assumes that the energy levels of the relativistic hydrogen atom are modified by terms proportional to $\alpha_0^2 \langle p_0^2 \rangle / M_\mathrm{Pl}^2 c^2$ (similar to (\ref{eqn:energy_levels_GUP}) for the non-relativistic case), 
deviations from the standard frequency of the 2$S$ - 1$S$ transition of the order $\sim \alpha_0^2 10^{-35}$ Hz are expected.
Comparing this with the result in 
\cite{Matveev2013}, we see that $\alpha_0 \lesssim 10^{18}$.
This further motivates a study of GUP for the relativistic hydrogen atom, 
in which the relativistic corrections should impose a much tighter bound on $\alpha_0$.
Similarly, for the Stern-Gerlach experiment, one can obtain an estimation of the parameter $\alpha_0$ directly comparing the relative error in the splitting $\delta z$ of an experiment with 
\eqref{eqn:cor_stern-gerlach}.
For example, for a relative error of 10\%, one obtains $\alpha_0 \lesssim 10^{14}$.
Notice that more accurate experiments would result in more stringent bounds.

There remain issues to be better understood, \emph{e.g.} the dependence of the angular momentum algebra on linear momentum, and some ambiguity in the CG coefficients.
Furthermore, we replaced the operator representing the GUP modification in some formulae by its expectation value.
While this suffice to estimate Planck scale effects, in the future we would like to study this further, to see if additional corrections result by retaining the operator forms.
The results presented here can be applied to look for QG signatures, \emph{e.g.} in spectroscopic as well as astrophysical observations.
Furthermore, assuming that the spin algebra also obeys similar modifications, they can also be applied to a number of quantum systems interacting with magnetic fields, in atomic and nuclear physics.
We hope to address these, as well as extensions of our work to relativistic QM \cite{Bosso}, in future publications.

\section*{Acknowledgments}
The authors thank E. C. Vagenas for discussions, and the Referees for useful comments.
This work was supported in part by the Natural Sciences and Engineering Research Council of Canada.

\begin{appendices}

\section{GUP modified angular momentum commutator} \label{apx:commutator_LiLj}

Consider the commutator between two components of the angular momentum
\begin{equation}
	[L_i,L_j] = \epsilon_{imn} \epsilon_{jrs} [q_m p_n, q_r p_s] = \epsilon_{imn} \epsilon_{irs} \{ q_m [ p_n, q_r ] p_s + q_r [q_m , p_s ] p_n \}~, \label{eqn:commutator_ang_mom}
\end{equation}
Using the GUP commutator in (\ref{eqn:GUP}), we obtain
\begin{multline}
		[L_i,L_j] = i \hbar \epsilon_{imn} \epsilon_{irs} \left \{ q_r p_n \left[ \delta_{ms} - \alpha \left(\delta_{ms} p + \frac{p_m p_s}{p} \right) + \alpha^2 (\delta_{ms} p^2 + 3 p_m p_s ) \right] + \right. \\
		\left. - q_m p_s \left[ \delta_{nr} - \alpha \left( \delta_{nr} p + \frac{p_n p_r}{p} \right) + \alpha^2 ( \delta_{nr} p^2 + 3 p_n p_r ) \right] \right\} = \\
		= i \hbar ( \epsilon_{mni} \epsilon_{mjr} q_r p_n - \epsilon_{nim} \epsilon_{nsj} q_m p_s ) ( 1 - \alpha p + \alpha^2 p^2) = \\
		 = i \hbar [ ( \delta_{ir} \delta_{nj} - \delta_{nr} \delta_{ij} ) q_r p_n - ( \delta_{is} \delta_{mj} - \delta_{ij} \delta_{ms} ) q_m p_s ] ( 1 - \alpha p + \alpha^2 p^2) 
		  = i \hbar \epsilon_{ijk} L_k ( 1 - \alpha p + \alpha^2 p^2)~,
\end{multline}

Consider now
\begin{equation}
	[L_i,p_l] = \epsilon_{ijk}(q_j p_k p_l - p_l q_j p_k) =
	i \hbar \epsilon_{ijk} \left[ \delta_{jl} p_k - \alpha \left( \delta_{jl} p p_k + \frac{p_j p_l p_k}{p} \right) + \alpha^2 (\delta_{jl} p^2 p_k + 3 p_j p_l p_k) \right]~,
\end{equation}
where we used the model in (\ref{eqn:GUP}).
We then have
\begin{equation}
	[L_i,p^2] = p_l[L_i,p_l] + [L_i,p_l]p_l = 0~. \label{eqn:commutator_Li_p2}
\end{equation}

To find the commutation relation between a component of the angular momentum and the magnitude of the linear momentum, we will first suppose that such a commutator depends only on the vector $\mathbf{p}$
\begin{equation}
	[L_i,p] = f(\mathbf{p})~.
\end{equation}
In this way, using the result just found we have
\begin{equation}
	0 = [L_i,p^2] = p[L_i,p] + [L_i,p]p = 2p[L_i,p]~,
\end{equation}
that means, for $p\not=0$,
\begin{equation}
	[L_i,p] = 0~. \label{eqn:commutator_Li_p}
\end{equation}

Finally we find
\begin{multline}
	[L^2,L_j] = L_i[L_i,L_j] + [L_i,L_j]L_i = i \hbar \epsilon_{ijk} [L_i L_k (1 - \alpha p + \alpha^2 p^2) + L_k (1 - \alpha p + \alpha^2 p^2) L_i] = \\
	= i \hbar \epsilon_{ijk} (L_i L_k + L_k L_i) (1 - \alpha p + \alpha^2 p^2) = 0~.
\end{multline}

\section{Clebsch-Gordan Coefficients}\label{apx:CG}
In this section, we will calculate several CG coefficients for different values of the total azimuthal and magnetic quantum numbers, L and M, referring these values to the maximum values $L_{\mathrm{max}} = l_1+l_2$ and $M_{\mathrm{max}} = L_{\mathrm{max}}$, where $l_1$ and $l_2$ are the azimuthal quantum numbers of the single systems.

\subsection{$L=L_{\mathrm{max}}$, $M=L_{\mathrm{max}}$}
For this case, the state represented by the total angular momentum can be related to just one of the states concerning the individual angular momenta, that is
\begin{equation}
	|L_\mathrm{max},L_\mathrm{max}\rangle = |l_1;l_2\rangle~.
\end{equation}
This means that the CG coefficient for this case is simply
\begin{equation}
	\langle l_1;l_2|L_\mathrm{max},L_\mathrm{max}\rangle = 1~. \label{eqn:CG_max_max}
\end{equation}

\subsection{$M=L_{\mathrm{max}}-1$}
Two states are possible
\begin{align}
		|l_1-1;l_2\rangle~,& & |l_1;l_2-1\rangle~.&
\end{align}

\subsubsection{$L = L_{\mathrm{max}}$}
Applying the lowering operator $L_- = l_{1,-} + l_{2,-}$ we find
\begin{multline}
	|L_\mathrm{max},L_\mathrm{max}-1\rangle \propto  L_-|L_\mathrm{max},L_\mathrm{max}\rangle = (l_{1,-} + l_{2,-})|l_1;l_2\rangle = \\
	= \hbar[(1- \langle \mathcal{C}_1 \rangle)\sqrt{2 l_1}|l_1-1;l_2\rangle + (1- \langle \mathcal{C}_2 \rangle ) \sqrt{2 l_2}|l_1;l_2-1\rangle]~,
\end{multline}
where we used the result in (\ref{eqn:CG_max_max}) and the relation (\ref{eqn:L+-}).
Since both these coefficients are positive (we are assuming that $\langle \mathcal{C} \rangle$ is smaller than 1), the Condon–Shortley phase convention is already fulfilled, we need just to normalize this combination since
\begin{equation}
	||L_-|L_\mathrm{max},L_\mathrm{max} -1 \rangle||^2 = 2\hbar^2 [(1- \langle \mathcal{C}_1 \rangle )^2 l_1 + (1- \langle \mathcal{C}_2 \rangle )^2 l_2]~.
\end{equation}
Thus, the two CG coefficient for this case are
\begin{subequations} \label{eqn:CG_0-1}
	\begin{align}
		\langle l_1-1;l_2|L_{\mathrm{max}},L_{\mathrm{max}}-1\rangle &= \frac{(1- \langle \mathcal{C}_1 \rangle )\sqrt{l_1}}{\sqrt{(1- \langle \mathcal{C}_1 \rangle )^2 l_1 + (1- \langle \mathcal{C}_2 \rangle )^2 l_2}}~, \\
		\langle l_1;l_2-1|L_{\mathrm{max}},L_{\mathrm{max}}-1\rangle &= \frac{(1- \langle \mathcal{C}_2 \rangle )\sqrt{l_2}}{\sqrt{(1- \langle \mathcal{C}_1 \rangle )^2 l_1 + (1- \langle \mathcal{C}_2 \rangle )^2 l_2}}~.
	\end{align}
\end{subequations}

\subsubsection{$L = L_{\mathrm{max}}-1$}
In this case, we will find the two CG coefficients for $|L_{\mathrm{max}}-1,L_{\mathrm{max}}-1\rangle$ applying the orthonormality condition between this state and $|L_{\mathrm{max}},L_{\mathrm{max}}-1\rangle$.
The state in this case can be written as a linear combination of $|l_1-1;l_2\rangle$ and $| l_1;l_2-1\rangle$
\begin{equation}
	|L_\mathrm{max}-1,L_\mathrm{max}-1\rangle = G_{10}|l_1-1;l_2\rangle + G_{01}| l_1;l_2-1\rangle~.
\end{equation}
From the orthogonality condition we find
\begin{equation}
	\langle L_\mathrm{max},L_\mathrm{max} -1 |L_\mathrm{max}-1,L_\mathrm{max}-1\rangle = \frac{(1- \langle \mathcal{C}_1 \rangle )\sqrt{l_1}}{\sqrt{(1- \langle \mathcal{C}_1 \rangle )^2 l_1 + (1- \langle \mathcal{C}_2 \rangle )^2 l_2}} G_{10} + \frac{(1- \langle \mathcal{C}_2 \rangle )\sqrt{l_2}}{\sqrt{(1- \langle \mathcal{C}_1 \rangle )^2 l_1 + (1- \langle \mathcal{C}_2 \rangle )^2 l_2}} G_{01} = 0~,
\end{equation}
obtaining
\begin{equation}
	G_{01} = - \frac{(1- \langle \mathcal{C}_1 \rangle )\sqrt{l_1}}{(1- \langle \mathcal{C}_2 \rangle )\sqrt{l_2}}G_{10}~.
\end{equation}
Normalizing the state
\begin{equation}
	|G_{10}|^2 \left[1 + \frac{(1- \langle \mathcal{C}_1 \rangle )^2 l_1}{(1- \langle \mathcal{C}_2 \rangle )^2 l_2} \right] = |G_{10}|^2 \frac{(1- \langle \mathcal{C}_1 \rangle )^2 l_1 + (1- \langle \mathcal{C}_2 \rangle )^2 l_2}{(1- \langle \mathcal{C}_2 \rangle )^2 l_2} = 1~,
\end{equation}
and imposing the Condon–Shortley phase convention
\begin{equation}
	\langle l_1; L_\mathrm{max}-1-l_1| L_\mathrm{max}-1,L_\mathrm{max}-1\rangle = \langle l_1; l_2-1| L_\mathrm{max}-1,L_\mathrm{max}-1\rangle \geq 0~,
\end{equation}
we have
\begin{subequations}
	\begin{align}
		\langle l_1-1;l_2|L_{\mathrm{max}}-1,L_{\mathrm{max}}-1\rangle &=  - \frac{(1- \langle \mathcal{C}_2 \rangle ) \sqrt{l_2}}{\sqrt{(1- \langle \mathcal{C}_1 \rangle )^2 l_1 + (1- \langle \mathcal{C}_2 \rangle )^2 l_2}}~, \\
		\langle l_1;l_2-1|L_{\mathrm{max}}-1,L_{\mathrm{max}}-1\rangle &= \frac{(1- \langle \mathcal{C}_1 \rangle ) \sqrt{l_1}}{\sqrt{(1- \langle \mathcal{C}_1 \rangle )^2 l_1 + (1- \langle \mathcal{C}_2 \rangle)^2 l_2}}.
	\end{align}
\end{subequations}

\subsection{$M=L_{\mathrm{max}}-2$}
In this case the three possible states are
\begin{align}
	|l_1,l_1-2;l_2,l_2\rangle~,& & |l_1,l_1-1;l_2,l_2-1\rangle~,& & |l_1,l_1;l_2,l_2-2\rangle~.&
\end{align}

\subsubsection{$L=L_{\mathrm{max}}$}
The CG coefficients for this case are found acting one time with $L_-$ on the state $|L_{\mathrm{max}},L_{\mathrm{max}}-1\rangle$
\begin{multline}
	L_- |L_{\mathrm{max}},L_{\mathrm{max}}-2\rangle \propto (l_{1,-} + l_{2,-})[\langle l_1 - 1;l_2|L_{\mathrm{max}},L_{\mathrm{max}}-1\rangle |l_1 - 1;l_2\rangle + \langle l_1;l_2 - 1|L_{\mathrm{max}},L_{\mathrm{max}}-1\rangle |l_1;l_2-1\rangle] = \\
	= \hbar \{\langle l_1 - 1;l_2|L_{\mathrm{max}},L_{\mathrm{max}}-1\rangle[(1- \langle \mathcal{C}_1 \rangle ) \sqrt{4l_1 - 2}|l_1 - 2;l_2\rangle + (1- \langle \mathcal{C}_2 \rangle )\sqrt{2l_2}|l_1-1;l_2-1\rangle] + \\
	+ \langle l_1;l_2 - 1|L_{\mathrm{max}},L_{\mathrm{max}}-1\rangle[(1- \langle \mathcal{C}_1 \rangle )\sqrt{2l_1}|l_1 - 1;l_2-1\rangle + (1- \langle \mathcal{C}_2 \rangle )\sqrt{4l_2 - 2}|l_1;l_2-2\rangle]\} = \\
	= \hbar\left\{\frac{\sqrt{2}(1- \langle \mathcal{C}_1 \rangle )^2\sqrt{l_1(2l_1-1)}}{\sqrt{(1- \langle \mathcal{C}_1 \rangle  )^2l_1 + (1- \langle \mathcal{C}_2 \rangle )^2l_2}}|l_1 - 2;l_2\rangle + \frac{2\sqrt{2}(1- \langle \mathcal{C}_1 \rangle )(1- \langle \mathcal{C}_2 \rangle )\sqrt{l_1l_2}}{\sqrt{(1-  \langle \mathcal{C}_1 \rangle )^2l_1 + (1- \langle \mathcal{C}_2 \rangle )^2l_2}}|l_1-1;l_2-1\rangle + \right.\\
	+ \left. \frac{\sqrt{2}(1- \langle \mathcal{C}_2 \rangle )^2\sqrt{l_2(2l_2-1)}}{\sqrt{(1- \langle \mathcal{C}_2 \rangle )^2l_2 + (1- \langle \mathcal{C}_1 \rangle )^2l_1}}|l_1;l_2-2\rangle \right\}~,
\end{multline}
where we used the relation (\ref{eqn:L+-}) and the coefficients in (\ref{eqn:CG_0-1}).
Normalizing this last result we find
\begin{subequations}
	\begin{align}
		\langle l_1 - 2;l_2| L_\mathrm{max}, L_\mathrm{max}-2\rangle &= \frac{(1- \langle \mathcal{C}_1 \rangle )^2\sqrt{l_1(2l_1-1)}}{\Omega_0}~, \\
		\langle l_1 - 1;l_2 - 1| L_\mathrm{max}, L_\mathrm{max}-2\rangle &= \frac{2(1- \langle \mathcal{C}_1 \rangle )(1- \langle \mathcal{C}_2 \rangle )\sqrt{l_1l_2}}{\Omega_0}~, \\
		\langle l_1;l_2 - 2| L_\mathrm{max}, L_\mathrm{max}-2\rangle &= \frac{(1- \langle \mathcal{C}_2 \rangle )^2\sqrt{l_2(2l_2-1)}}{\Omega_0}~,
	\end{align}
\end{subequations}
where
\begin{equation}
	\Omega_0 = \sqrt{(1- \langle \mathcal{C}_1 \rangle )^4l_1(2l_1-1) + 4(1- \langle \mathcal{C}_1 \rangle )^2(1- \langle \mathcal{C}_2 \rangle )^2l_1l_2 + (1- \langle \mathcal{C}_2 \rangle  )^4l_2(2l_2-1)}~.
\end{equation}
Since these coefficients are all positive, the phase convention is already fulfilled.

\subsubsection{$L=L_{\mathrm{max}} - 1$}

To find the CG coefficients for this case we apply $L_-$ on the state $|L_{\mathrm{max}} - 1, L_{\mathrm{max}} - 1\rangle$
\begin{multline}
	L_-|L_{\mathrm{max}} - 1, L_{\mathrm{max}} - 1\rangle = \\
	= (l_{1,-} + l_{2,-})[\langle l_1 - 1;l_2|L_{\mathrm{max}}-1,L_{\mathrm{max}}-1\rangle |l_1 - 1;l_2\rangle + \langle l_1;l_2 - 1|L_{\mathrm{max}}-1,L_{\mathrm{max}}-1\rangle |l_1;l_2-1\rangle] = \\
	= \hbar \langle l_1 - 1;l_2|L_{\mathrm{max}}-1,L_{\mathrm{max}}-1\rangle[(1-\langle \mathcal{C}_1 \rangle)\sqrt{4l_1 -2}|l_1 - 2;l_2\rangle + (1-\langle \mathcal{C}_2 \rangle)\sqrt{2l_2}|l_1 - 1;l_2-1\rangle] + \\
	+ \hbar \langle l_1;l_2-1|L_{\mathrm{max}}-1,L_{\mathrm{max}}-1\rangle[(1-\langle \mathcal{C}_1 \rangle)\sqrt{2l_1}|l_1 - 1;l_2-1\rangle + (1-\langle \mathcal{C}_2 \rangle)\sqrt{4l_2 - 2}|l_1;l_2-2\rangle] \propto \\
	\propto - \hbar (1-\langle \mathcal{C}_1 \rangle)(1-\langle \mathcal{C}_2 \rangle)\sqrt{4l_1 -2}\sqrt{l_2}|l_1 - 2;l_2\rangle + \hbar [(1-\langle \mathcal{C}_1 \rangle)^2 \sqrt{2} l_1 - (1-\langle \mathcal{C}_2 \rangle)^2 \sqrt{2} l_2]|l_1 - 1;l_2-1\rangle + \\
	+ \hbar(1-\langle \mathcal{C}_1 \rangle)(1-\langle \mathcal{C}_2 \rangle) \sqrt{l_1}\sqrt{4l_2 - 2}|l_1 - 1;l_2-2\rangle~.
\end{multline}
Normalizing and using the Condon-Shortley convention we thus find
\begin{subequations}
	\begin{align}
		\langle l_1 - 2;l_2| L_\mathrm{max}-1, L_\mathrm{max}-2\rangle &= \frac{(1-\langle \mathcal{C}_1 \rangle)(1-\langle \mathcal{C}_2 \rangle)\sqrt{4l_1 -2}\sqrt{l_2}}{\Omega_1}~, \\
		\langle l_1 - 1;l_2 - 1| L_\mathrm{max} -1, L_\mathrm{max}-2\rangle &= - \frac{(1-\langle \mathcal{C}_1 \rangle)^2 \sqrt{2} l_1 - (1-\langle \mathcal{C}_2 \rangle)^2 \sqrt{2} l_2}{\Omega_1}~, \\
		\langle l_1;l_2 - 2| L_\mathrm{max} - 1, L_\mathrm{max}-2\rangle &= - \frac{(1-\langle \mathcal{C}_1 \rangle)(1-\langle \mathcal{C}_2 \rangle) \sqrt{l_1}\sqrt{4l_2 - 2}}{\Omega_1}~,
	\end{align}
\end{subequations}
with
\begin{equation}
	\Omega_1 = \sqrt{2(1-\langle \mathcal{C}_1 \rangle)^4 l_1^2 + 2(1-\langle \mathcal{C}_1 \rangle)^2(1-\langle \mathcal{C}_2 \rangle)^2(2l_1l_2 - l_1 - l_2) + 2(1-\langle \mathcal{C}_2 \rangle)^4l_2^2}~.
\end{equation}

\subsubsection{$L=L_{\mathrm{max}} - 2$}
As first step, let us define the CG coefficients for this case in the following way
\begin{equation}
	|L_{\mathrm{max}} - 2, L_{\mathrm{max}} - 2\rangle = G_{20}|l_1-2,l_2\rangle + G_{11}|l_1-1,l_2-1\rangle + G_{02}|l_1,l_2-2\rangle~.
\end{equation}
Considering the orthogonality between the states $|L_{\mathrm{max}} - 2, L_{\mathrm{max}} - 2\rangle$ and $|L_{\mathrm{max}} - 1,L_{\mathrm{max}} - 2\rangle$
\begin{multline}
	-G_{20} (1-\langle \mathcal{C}_1 \rangle)(1-\langle \mathcal{C}_2 \rangle)\sqrt{4l_1 -2}\sqrt{l_2}  + G_{11}[(1-\langle \mathcal{C}_1 \rangle)^2 \sqrt{2} l_1 - (1-\langle \mathcal{C}_2 \rangle)^2 \sqrt{2} l_2] + \\
	+ G_{02}(1-\langle \mathcal{C}_1 \rangle)(1-\langle \mathcal{C}_2 \rangle) \sqrt{l_1}\sqrt{4l_2 - 2} = 0~, \label{eqn:-2-2_step1}
\end{multline}
and the orthogonality between the first state and $|L_{\mathrm{max}}, L_{\mathrm{max}} - 2\rangle$
\begin{equation}
	G_{20} (1-\langle \mathcal{C}_1 \rangle)^2\sqrt{l_1(2l_1-1)}  + 2G_{11}(1-\langle \mathcal{C}_1 \rangle)(1-\langle \mathcal{C}_2 \rangle)\sqrt{l_1l_2}  + G_{02}(1-\langle \mathcal{C}_2 \rangle)^2\sqrt{l_2(2l_2-1)} = 0~,
\end{equation}
whence
\begin{equation}
	G_{11} = - G_{20} \frac{1-\langle \mathcal{C}_1 \rangle}{1-\langle \mathcal{C}_2 \rangle}\frac{\sqrt{2l_1-1}}{2\sqrt{l_2}} - G_{02} \frac{1-\langle \mathcal{C}_2 \rangle}{1-\langle \mathcal{C}_1 \rangle}\frac{\sqrt{2l_2-1}}{2\sqrt{l_1}}~. \label{eqn:-2-2_step2}
\end{equation}
Inserting this last result in (\ref{eqn:-2-2_step1}) we find
\begin{multline}
	- G_{20}\frac{1-\langle \mathcal{C}_1 \rangle}{1-\langle \mathcal{C}_2 \rangle} \frac{\sqrt{2l_1-1}}{\sqrt{2l_2}} \left[ (1-\langle \mathcal{C}_2 \rangle)^2 l_2 + (1-\langle \mathcal{C}_1 \rangle)^2 l_1\right] + \\
	+ G_{02}\frac{1-\langle \mathcal{C}_2 \rangle}{1-\langle \mathcal{C}_1 \rangle} \frac{\sqrt{2l_2-1}}{\sqrt{2l_1}} \left[ (1-\langle \mathcal{C}_1 \rangle)^2 l_1 + (1-\langle \mathcal{C}_2 \rangle)^2 l_2 \right] = 0 \Rightarrow\\
	\Rightarrow G_{20} = G_{02}\frac{\sqrt{l_2(2l_2-1)}}{\sqrt{l_1(2l_1-1)}}~,
\end{multline}
whence, using this result in (\ref{eqn:-2-2_step2}) we obtain
\begin{equation}
	G_{11} = - G_{02} \frac{\sqrt{2l_2-1}}{2\sqrt{l_1}} \frac{(1-\langle \mathcal{C}_1 \rangle)^2 + (1-\langle \mathcal{C}_2 \rangle)^2}{(1-\langle \mathcal{C}_1 \rangle)(1-\langle \mathcal{C}_2 \rangle)}~.
\end{equation}
Imposing the normalization condition
\begin{equation}
		|G_{02}|^2\frac{l_2(2l_2 - 1)}{l_1(2l_1 - 1)} + |G_{02}|^2 \frac{2l_2-1}{4l_1} \frac{(1-\langle \mathcal{C}_1 \rangle)^4 + (1-\langle \mathcal{C}_2 \rangle)^4 + 2(1-\langle \mathcal{C}_1 \rangle)^2(1-\langle \mathcal{C}_2 \rangle)^2}{(1-\langle \mathcal{C}_1 \rangle)^2(1-\langle \mathcal{C}_2 \rangle)^2} + |G_{02}|^2 = 1
\end{equation}
and the Condon-Shortley phase convention we have
\begin{subequations}
	\begin{align}
		\langle l_1;l_2 - 2| L_\mathrm{max} - 2, L_\mathrm{max}-2\rangle =& \frac{2 \sqrt{l_2(2l_2-1)} (1-\langle \mathcal{C}_1 \rangle)(1-\langle \mathcal{C}_2 \rangle)}{\Omega_2}~,\\
		\langle l_1;l_2 - 2| L_\mathrm{max} - 2, L_\mathrm{max}-2\rangle =& - \frac{\sqrt{(2l_1-1)(2l_2-1)}[(1-\langle \mathcal{C}_1 \rangle)^2 + (1-\langle \mathcal{C}_2 \rangle)^2]}{\Omega_2}~,\\
		\langle l_1;l_2 - 2| L_\mathrm{max} - 2, L_\mathrm{max}-2\rangle =& \frac{2 \sqrt{l_1(2l_1-1)} (1-\langle \mathcal{C}_1 \rangle)(1-\langle \mathcal{C}_2 \rangle)}{\Omega_2}~,
	\end{align}
\end{subequations}
where
\begin{multline}
	\Omega_2 = \{2(1-\langle \mathcal{C}_1 \rangle)^2(1-\langle \mathcal{C}_2 \rangle)^2[2l_2(2l_2 - 1) + 2l_1(2l_1 - 1) + (2l_1-1)(2l_2-1)] + \\
	+ (2l_1-1)(2l_2-1) [(1-\langle \mathcal{C}_1 \rangle)^4 + (1-\langle \mathcal{C}_2 \rangle)^4]\}^{1/2}~.
\end{multline}

\end{appendices}

\end{document}